%% file: main.tex
\newif\ifANONYMOUS
\ANONYMOUSfalse

\documentclass[acmsmall,screen,nonacm]{acmart}

\newif\ifSUPPLEMENT
\SUPPLEMENTtrue

\newif\ifDEBUG
\DEBUGfalse

\usepackage{soul}
\usepackage{algorithmic}
\usepackage{graphicx}
\usepackage{textcomp}
\usepackage{xcolor}
\usepackage{acronym}
\usepackage{xtab}
\usepackage{url}
\usepackage{hyperref}
\usepackage{subcaption}
\usepackage[font=small,skip=2pt]{caption} %
\usepackage{picins}
\usepackage{multirow}
\usepackage{pgfplots}
\usepackage{enumitem}
\usepgfplotslibrary{statistics}
\pgfplotsset{width=0.48\textwidth,compat=1.18}
\usepackage{pgfplotstable}
\usepackage{xspace}
\usepackage{fontawesome}
\usepackage{url}
\usepackage{tablefootnote}
\usepackage{xcolor,colortbl} %
 \usepackage{booktabs}
\usepackage{wrapfig}

\usepackage{cleveref}
    \crefformat{section}{\S#2#1#3}
    \crefname{figure}{Figure}{Figures}
    \crefname{appendix}{Appendix}{Appendices}
    \crefname{table}{Table}{Tables}
    \crefname{algorithm}{Algorithm}{Algorithms}
    \crefname{listing}{Listing}{Listings}
    \crefname{theorem}{Theorem}{Theorems}
    \crefname{thm}{Theorem}{Theorems}
    \crefname{lemma}{Lemma}{Lemmata}
    \crefname{equation}{Eqt.}{Eqts.}
    \crefformat{Grammar}{Grammar #1}

\usepackage[newfloat=true,cachedir=minted-cache,frozencache]{minted}
\setminted[]{
baselinestretch=1.1,
fontsize=\footnotesize,
fontfamily=tt,
numbersep=5pt,
breaklines,
escapeinside=@@,
breakautoindent=true,
breakafter=\space/(,
autogobble,
}

\input{macros}

\input{data/data}

\AtBeginDocument{%
  }

\setcopyright{cc}
\setcctype{by}
\acmYear{2025}

\begin{document}
\newcommand{\mytitle}{}
\renewcommand{\mytitle}{Finding \numalldefects{} Defects in \numrepos Projects: An Experience Report on Applying CodeQL to Open-Source Embedded Software (Experience Paper)}

\title[Finding \numalldefects{} Defects in \numrepos Projects: An Experience Report on Applying CodeQL \ldots]{\mytitle%
\ifSUPPLEMENT
~-- Extended Report
\fi
}
\titlenote{This is the extended version of:
Mingjie Shen, Akul Abhilash Pillai, Brian A. Yuan, James C. Davis, and Aravind Machiry. 2025. Finding 709 Defects in 258 Projects: An Experience Report on Applying CodeQL to Open-Source Embedded Software (Experience Paper). \textit{Proc. ACM Softw. Eng.} 2, ISSTA, Article ISSTA048 (July 2025), 24 pages. \url{https://doi.org/10.1145/3728923}
}

\pgfplotsset{
    only if/.style args={entry of #1 is #2}{
        /pgfplots/boxplot/data filter/.code={
            \edef\tempa{\thisrow{#1}}
            \edef\tempb{#2}
            \ifx\tempa\tempb
            \else
            \def\pgfmathresult{}
            \fi
        }
    }
}

\ifANONYMOUS
\else
\author{Mingjie Shen}
\orcid{0009-0003-4393-6992}
\affiliation{%
  \institution{Purdue University}
  \city{West Lafayette}
  \country{USA}
}
\email{shen497@purdue.edu}

\author{Akul Abhilash Pillai}
\orcid{0000-0002-0588-6440}
\affiliation{%
  \institution{Purdue University}
  \city{West Lafayette}
  \country{USA}
}
\email{pillai23@purdue.edu}

\author{Brian A. Yuan}
\orcid{0009-0007-5723-9070}
\affiliation{%
  \institution{Purdue University}
  \city{West Lafayette}
  \country{USA}
}
\email{bryuan@student.ethz.ch}
\authornote{Work performed while at Purdue University.}

\author{James C. Davis}
\orcid{0000-0003-2495-686X}
\affiliation{%
  \institution{Purdue University}
  \city{West Lafayette}
  \country{USA}
}
\email{davisjam@purdue.edu}

\author{Aravind Machiry}
\orcid{0000-0001-5124-6818}
\affiliation{%
  \institution{Purdue University}
  \city{West Lafayette}
  \country{USA}
}
\email{amachiry@purdue.edu}

\fi

\ifANONYMOUS
\renewcommand{\shortauthors}{Anonymous et al.}
\else
\renewcommand{\shortauthors}{M. Shen, A. Pillai, B. Yuan, J. Davis, and A. Machiry}
\fi

\begin{abstract}
Embedded software is deployed in billions of devices worldwide, including in safety-sensitive systems like medical devices and autonomous vehicles.
Defects in embedded software can have severe consequences.
Many embedded software products incorporate~\acf{EMBOSS}, so it is important for~\ac{EMBOSS} engineers to use appropriate mechanisms to avoid defects.
One of the common security practices is to use \ac{SAST} tools, which help identify commonly occurring vulnerabilities.
Existing research related to \ac{SAST} tools focuses mainly on regular (or non-embedded) software. 
There is a lack of knowledge about the use of \ac{SAST} tools in embedded software.
Furthermore, embedded software greatly differs from regular software in terms of semantics, software organization, coding practices, and build setup.
All of these factors influence \ac{SAST} tools and could potentially affect their usage.

In this experience paper, we report on a large-scale empirical study of \ac{SAST} in \ac{EMBOSS} repositories.
We collected a corpus of~\numrepos{} of the most popular \ac{EMBOSS} projects, %
and then measured their \textit{use} of~\ac{SAST} tools via program analysis and a survey (N=25) of their developers. Advanced \ac{SAST} tools are rarely used -- only 3\% of projects go beyond trivial compiler analyses. Developers cited the perception of ineffectiveness and false positives as reasons for limited adoption.
Motivated by this deficit, we applied the \ac{SOTA} \codeql{}~\ac{SAST} tool and measured its ease of use and actual effectiveness.
Across the~\numrepos{} projects, \codeql{} reported~\numalldefects{} true defects with a false positive rate of~\codeqlfpperc.
There were~\numsecuritydefects{} (\numsecuritydefectsperc{}) likely security vulnerabilities, including in major projects maintained by Microsoft, Amazon, and the Apache Foundation.
\ac{EMBOSS} engineers have confirmed ~\numallconfirmeddefects{} (\numallconfirmeddefectsperc{}) of these defects, mainly by accepting our pull requests.
Two CVEs were issued.
Based on these results, we proposed pull requests to include our workflows as part of~\ac{EMBOSS}~\ac{CI} pipelines, \nummergedworkflows{} (\percmergedwfactiverepos of active repositories) of these are already merged.
In summary, we urge~\ac{EMBOSS} engineers to adopt the current generation of~\ac{SAST} tools, which offer low false positive rates and are effective at finding security-relevant defects.
\end{abstract}

\ccsdesc[500]{Computer systems organization~Embedded software}
\ccsdesc[500]{Security and privacy~Software and application security}
\ccsdesc[500]{Security and privacy~Systems security}

\keywords{Empirical Software Engineering, Static Application Security Testing (SAST)}

\maketitle

\section{Introduction}

\label{sec:intro}

Societies rely on embedded systems and IoT devices in our
  transportation~\cite{al2020intelligence},
  traffic management~\cite{soni2017review},
  resource distribution~\cite{prapti2022internet,o2013industrial},
  homes~\cite{Alrawi2019SoKDeployments},
  and in many other ways~\cite{al-garadi_survey_2020}.
The~\ac{EmS} that enables these devices must be free of vulnerabilities.
Such vulnerabilities have far-reaching consequences~\cite{antonakakis2017understanding, writer_5_2020, margolis2017depth, 8688434, bonaventura2023smart} due to the pervasive and interconnected nature of embedded devices.
Additionally, \ac{OSS} plays an important role in~\ac{EmS} development~\cite{10.1145/2737182.2737190, lundellPractitionerPerceptionsOpen2011, amiri2017survey}. For instance, FreeRTOS~\cite{noauthor_freertos_nodate} and Zephyr~\cite{the_linux_foundation_zephyr_2023}, two of the most popular and industry-endorsed~\acp{RTOS}, are open-source.
Previous studies~\cite{rustforembedded, emnesttest} show that \acf{EMBOSS} are riddled with security vulnerabilities, specifically memory safety issues.

Several static and dynamic analysis-based tools exist for vulnerability detection.
Dynamic analyses, such as fuzzing~\cite{manes2019art}, are known to be effective at precise vulnerability detection.
However, applying these techniques to embedded systems is challenging~\cite{muench2018you,white_making_2011} because of their close interaction with hardware and its diversity.
Static analysis techniques, specifically \ac{SAST} tools, are best suited as they do not need to execute the embedded software or \ac{EMBOSS}.
On the other hand, most existing works~\cite{julietusedtoevalsasttools, santoso2021implementation, phan2023challenges, imtiaz2019synopsis} on evaluating \ac{SAST} tools focus on traditional (\ie{} non-embedded) software.
However, embedded software differs from traditional software in organization, architecture, build system, and toolchains~\cite{white_making_2011}.
\emph{What would be the effectiveness of \ac{SAST} tools on \ac{EMBOSS}?}

In this experience paper, we present the first empirical study on the use of~\ac{SAST} tools to detect security vulnerabilities in~\ac{EMBOSS}.
For our study, we curated a corpus of~\numrepos popular~\ac{EMBOSS} projects from GitHub.
We used this corpus for the three phases of our investigation.

\textbf{(1) Measuring the use of SAST in EMBOSS:}
First, we combined automated analysis of~\ac{CI} workflows from the corpus and a survey of the project developers to understand the prevalence of~\ac{SAST} usage.
We found that only~\numprojectsast{} (\numprojectsastperc{}) projects use explicit~\ac{SAST} tools as part of their~\ac{CI} workflows.
Developers of the remaining projects are aware of~\ac{SAST} tools but do not use them on~\ac{EMBOSS} projects as they believe that the effectiveness of \ac{SAST} tools on their repositories is low.
It is unclear whether this belief is accurate.

\textbf{(2) Selecting and Configuring a SAST Tool:}
Next, to fill this knowledge gap, we aim to understand the effectiveness of \ac{SAST} on \ac{EMBOSS}.
We conducted a preliminary analysis and found that the~\codeql{} was the most effective available \ac{SAST} tool.
First, the default setup of \codeql{} (that works well for traditional software) failed on \ac{EMBOSS} repositories.
Furthermore, the default analysis resulted in a lot of false positives.
We tackled this by manually (with minimal engineering effort) creating~\ac{CI} workflows enabling the execution of~\codeql{} on~\ac{EMBOSS} repositories.
 Second, the default analyses queries of \codeql{} resulted in a lot of false positives. We tackled this by filtering out certain queries and modifying relevant queries.

\textbf{(3) Measuring the effectiveness of SAST in EMBOSS:}
We executed our~\ac{CI} workflows with modified \codeql{} queries and found a total of~\numalldefects{} defects, with~\numsecuritydefects{} (\numsecuritydefectsperc{}) being security vulnerabilities.
On a per-report basis,~\codeql exhibits a false positive rate of \codeqlfpperc, but this is due to a few outlier rules and projects.
For most studied repositories, the false positive rates were low.
We reported~\numdefectsdisclosed of defects we found.
Developers have already confirmed~\numallconfirmeddefects{} (\numallconfirmeddefectsperc{}) of these defects, mainly by accepting our patches.
We also raised pull requests to~\numcodeqlworkflowpullquests{}~\ac{EMBOSS}, integrating our manually created workflows (enabling running~\codeql{}) into their CI pipeline, out of which~\nummergedworkflows{} (\percmergedwfactiverepos{} (Active) and \nummergedworkflowsperc{} (Total)) are already accepted.
We hope that our findings:
  (1) provide evidence of the effectiveness of \ac{SAST} tool on \ac{EMBOSS} repositories;
  (2) encourage \ac{EMBOSS} developers to adopt \ac{SAST} tools;
  and
  (3) motivate researchers to work on techniques to automatically integrate \ac{SAST} tools in \ac{CI} workflows.

In summary, this experience report contributes:
\begin{itemize}[leftmargin=*]
\item~\textbf{(Empirical Study)} We presented the first study on the prevalence, challenges, and effectiveness of using \ac{SAST} tools in \ac{EMBOSS}, via automated and manual analysis and a developer study.
\item~\textbf{(Lessons Learned)} 
We summarize our experience in four lessons learned (\sect{sec:lessonslearned}), capturing our experiences using a \ac{SOTA} \ac{SAST} tool, finding hundreds of defects, reporting them, and integrating the \ac{SAST} tool in the \ac{CI} pipeline of \ac{EMBOSS} repositories.
\item~\textbf{(Dataset)} We curated and categorized a list of~\numrepos{} major~\ac{EMBOSS} projects, accompanied by GitHub workflows for compilation to permit the execution of static and dynamic analysis tools. This is the first large-scale embedded software dataset with compilation infrastructure. %

\item~\textbf{(Impact)} Using off-the-shelf CodeQL queries on these workflows, %
we identified a total of~\numalldefects{} defects (\numsecuritydefects{} (\numsecuritydefectsperc{}) security vulnerabilities) across this dataset, including projects maintained by the Apache Foundation, Microsoft, and Amazon.
We reported~\numdefectsdisclosed of these defects, of which %
developers confirmed~\numallconfirmeddefects{} (\numallconfirmeddefectsperc{}) of them. %
We also raised pull requests to~\numcodeqlworkflowpullquests{} projects to integrate~\codeql workflows in their~\ac{CI} pipelines, of which~\nummergedworkflows{} are accepted.

\end{itemize}

\noindent

\emph{Significance for software engineering:}
Empirical software security research has a substantial body of knowledge on open-source software, but has focused on IT or general-purpose software.
We report on a large-scale experience of applying static analysis to open-source embedded software. %
Across \numrepos \ac{EMBOSS} repositories, the \codeql \ac{SAST} tool finds hundreds of defects with low false positive rates in the majority of repositories.
Motivated by this knowledge, we recommend that \ac{EMBOSS} software developers use this tool to easily improve software quality.

\section{Background} \label{sec:Background}

Here we define \acf{EMBOSS}
  and
  Dynamic and \acf{SAST}. %

\subsection{Open-Source Embedded Software (EMBOSS)} \label{sec:Background-EMBOSS}

\subsubsection{Definition of Embedded Software and EMBOSS} \label{sec:Background-EMBOSS-Definition}
Embedded software is designed to run on embedded systems, ranging from industrial controllers~\cite{bhamare2020cybersecurity} to IoT devices with resource-constrained microcontrollers~\cite{Alrawi2019SoKDeployments}.

\acf{OSS} is an essential part of the software supply chain of embedded software applications. %
A considerable proportion of software products incorporate open-source software in order to reduce costs and develop more competitive products~\cite{amiri2017survey}.
Application developers re-use many kinds of EMBOSS, but a particularly common dependency is on specialized~\acfp{RTOS} designed for reduced-resource environments (\eg real-time scheduling, low power consumption, low memory overhead).
According to \code{osrtos.com}, there are 31 different~\acp{RTOS}~\cite{osrtosweb}, with the majority (26) of them being open-source.
Examples of~\acp{RTOS} include RIOT, Contiki, FreeRTOS, and Azure RTOS.

\subsubsection{Measuring Project Importance} \label{sec:Background-EMBOSS-OSSFCriticalityScore}
A common way to measure the importance of an open-source project is the~\ac{OSSF} criticality score~\cite{ossfcriticalityscore}.
This score is used by security analysts to triage security vulnerabilities when studying a large number of projects~\cite{mawanted, 10.1145/3543873.3587336}.
A project's importance is a number between 0 and 1 based on attributes including its popularity, dependents, and level of activity.
Ranges correspond to qualitative labels:
  0.0-0.2 is considered low criticality, 
  0.2-0.4 is medium,
  0.4-0.6 is high,
  0.6-0.9 is critical,
  and
  above 0.9 is extremely critical.
For examples,
  the RTOS contiki-os has a criticality score of 0.51 (high),
  the RTOS Zephyr's score is 0.81 (critical),
  and
  the Node.js runtime's score is 0.99 (extremely critical).

\subsection{Static Application Security Testing (SAST)} \label{sec:Background-SAST}

\subsubsection{SAST vs. DAST in Embedded Software} \label{sec:Background-AST-EMBOSS}
In software security analysis, both static (SAST) and dynamic (DAST) application security testing are necessary.

In the context of embedded systems, dynamic analysis (\eg{} fuzzing) is more costly and sometimes infeasible when compared to static analysis.
Embedded software is coupled to hardware~\cite{muench2018you}, \eg using hardware-specific interfaces and custom instruction sets.
Executing it on custom hardware needs an emulator (support may be lacking~\cite{fasano2021sok}) or physical boards (resulting in unscalable testing).
\acf{SAST} tools do not require execution, making them attractive to use on embedded software.

\subsubsection{Landscape of SAST Tools} %
There are many open-source and commercial~\ac{SAST} tools.
The open-source tools vary in the underlying techniques and corresponding guarantees.
There are high-assurance tools, such as IKOS~\cite{brat2014ikos}, that use abstract interpretation and provide soundness guarantees.
However, these tools must be properly configured with suitable abstract domains to avoid false positives -- a cumbersome process requiring a formal background.
On the other hand, there are best-effort pattern-based tools, such as \code{cppcheck}~\cite{marjamaki2013cppcheck} and \code{flawfinder}~\cite{wheeler2006flawfinder}, which can be readily used but do not provide any guarantees.
Several works~\cite{gentsch2020evaluation, chatzieleftheriou2011test, moerman2018evaluating, lu2018evaluating} evaluate these tools on non-embedded software and show that they vary in precision, recall, and usability.
There are also many commercial~\ac{SAST} tools.
Coverity is considered \ac{SOTA} and allows developers to customize the tool to reduce false positives~\cite{imtiaz_how_2019}, but its license forbids evaluation in research papers.
Other notable tools include Fortify~\cite{noauthor_fortify_nodate}, Checkmarx~\cite{noauthor_checkmarx_nodate}, and Veracode~\cite{noauthor_veracode_nodate}.

\codeql{} is a \ac{SOTA}~\cite{lipp_empirical_2022} open-source~\ac{SAST} tool.
\codeql{} was released in 2016 by GitHub and is maintained by Microsoft.
\codeql{} represents code as a relational database and uses relational queries to find defects in the given codebase.
It has several static analysis capabilities, such as control flow analysis, data flow analysis, and taint tracking to detect security issues~\cite{avgustinov2016ql}.
Furthermore,~\codeql{} has built-in queries for common security issues (\ie{}~\acp{CWE}).
Security analysts and developers have used~\codeql{} to find thousands of security vulnerabilities in large and well-tested codebases including the Linux kernel~\cite{codeqlwof, codeqlwof2, codeqlwof3}.
Since \codeql{} is free to use on open-source codebases and its queries are open-source, it is a popular SAST tool within the open-source community.

\subsubsection{How SAST is Applied in Modern OSS} \label{sec:Background-CI}

\acf{CI} pipelines~\cite{humble2010continuous} have become ubiquitous in the modern software development lifecycle.
They automate various software development processes, such as building, testing, and deploying code.
By this means, software development has shifted towards the continuous (or near-continuous~\cite{ananthanarayanan2019keeping}) integration of changes, allowing deployment at more rapid intervals~\cite{forsgren2018accelerate}.
SAST and DAST tools are often applied as part of a~\ac{CI} pipeline~\cite{bajpai2022secure,mangla2023securing,mansfield2018devops}, reflecting the ``shift left'' trend to assess security throughout the engineering process rather than at fixed intervals.

On GitHub, the main open-source software platform, there are several options for CI frameworks~\cite{decan2022use},
  \eg TravisCI~\cite{travicci},
  CircleCI~\cite{circleci},
  and GitHub Actions~\cite{chandrasekara2021introduction}.
The most popular is GitHub Actions because of its close integration with GitHub's platform~\cite{9825792}.
The GitHub CI is structured as a set of \textit{workflows} associated with events.
Each workflow is comprised of one or more \textit{Actions}.
\apdx{apdx:workflowbackground} provides more detail about GitHub workflows.

\section{Motivation}
\label{sec:RQs-Motivation}

Many works~\cite{julietusedtoevalsasttools, santoso2021implementation, phan2023challenges, imtiaz2019synopsis} emphasize the importance of using~\ac{SAST} tools on software projects, especially in unsafe languages such as C/C++ (which most~\ac{EMBOSS} repositories use).
Cybersecurity and government organizations~\cite{cernsecurity, gsasecurity} also recommend the use of~\ac{SAST}.
But no study measures the prevalence or benefits of~\ac{SAST} in~\ac{EMBOSS}.

Existing studies~\cite{gentsch2020evaluation, chatzieleftheriou2011test, moerman2018evaluating, lu2018evaluating, 8531281} evaluate the effectiveness of~\ac{SAST} tools on non-embedded software.
Embedded software differs from traditional software in organization, architecture, build system, and toolchains~\cite{white_making_2011}.
\ac{EMBOSS} follows a layered organization where each layer exposes fixed functionalities to the one above and relies on those below~\cite{10.1145/3589610.3596271}. To enhance flexibility, inter-layer communication uses function pointers, leading to indirect control flow transfers -- a common cause of static analysis imprecision.
Additionally, \ac{EMBOSS} employs an event-driven architecture with handlers triggered by specific events (\eg interrupts)~\cite{samek2008practical}. These handlers communicate via global objects, creating asynchronous control flows with global pointer manipulations, which challenge flow-based static analysis.
Furthermore, \ac{EMBOSS} relies on diverse, non-standard build systems~\cite{10.1145/3589610.3596271} and exotic compilers (\eg~\code{avr-gcc}), posing engineering difficulty in applying compilation-based \ac{SAST} tools.
Therefore, it is unclear how challenging it is to use existing~\ac{SAST} tools and how effective they are on~\ac{EMBOSS}.

\section{Prevalence Study}
Given the potential benefits of \ac{SAST} tools, we first study the prevalence of their usage in \ac{EMBOSS}.
We curate a \textit{corpus} of major~\ac{EMBOSS} from GitHub (\cref{sec:EMBOSSCorpus}) and other well-known sources.

\subsection{Corpus of Major EMBOSS Projects} \label{sec:EMBOSSCorpus}
\label{sec:EMBOSSCorpus-Approach}

{
\begin{table*}[ht]
\small
\caption{
  Summary of repositories in our~\ac{EMBOSS} dataset, grouped by project categories.
  SLOC calculated with \code{cloc}~\cite{adanial_cloc}.
  Criticality with the OSSF tool~\cite{ossfcriticalityscore}.
  Data is as of July, 2023.
  The \emph{Total} row gives medians across corpus, not by category.
  Medians are rounded to the nearest integer.
  }
\label{tab:EMBOSSCorpus-Summary}
\centering
\small
\begin{tabular}{lrlrrr}

\toprule
& & & \textbf{Median}& \textbf{Median}& \textbf{Median}\\
\textbf{Category} & \textbf{\# Repos} 
& \textbf{Example Repo} & \textbf{GH stars} & \textbf{SLOC}  & \textbf{Crit. Score}\\ \midrule 
Hardware Access Library (HAL)   & \numhardware              & grbl                 & 304& 98,502   & 0.44\\
Device Drivers (DD)            & \numdevicedriver          & TinyUSB              & 452& 20,078   & 0.41\\
Network (NET)                   & \numnetworks              & contik-ng            & 314& 36,345   & 0.46\\
Database Access Libraries (DAL) & \numdbaccess              & tiny SQL             & 659& 26,977   & 0.39\\
File Systems (FS)              & \numfilesys               & littlefs             & 401& 11,195   & 0.49\\
Parsing Utilities (PAR)         & \numparsing               & mjson               & 314& 2,547  & 0.41\\
Language Support (LS)          & \numlanguagesupport       & micropython          & 479& 33,389   & 0.42\\
UI Utilities (UI)              & \numUI                    & flutterpi            & 584& 56,712 & 0.46\\
Embedded Applications (APP)     & \numembedapp              & Infinitime           & 508& 22,662 & 0.39\\
Operating Systems (OS)           & \numOS                    & FreeRTOS              & 728& 409,668& 0.47\\ 
Memory Management Library (MML) & \nummemmanage             & tinyobjloader-c      & 242& 6,206& 0.34\\
Other General Purpose Library \\ \; \; for Embedded Use (GPL)
                          & \numgenlib                & tinyprintf           & 391& 12,742 & 0.35\\
Other (OT)                     & \numother                 &                      & 368& 94,805   & 0.43\\
\midrule
\textbf{Total}                     & \numrepos & & 406 & 33,545 &0.43                   \\
\bottomrule
\end{tabular}
\end{table*}
}

\begin{wrapfigure}{R}{0.5\textwidth}
 \centering
 \includegraphics[width=0.4\textwidth]{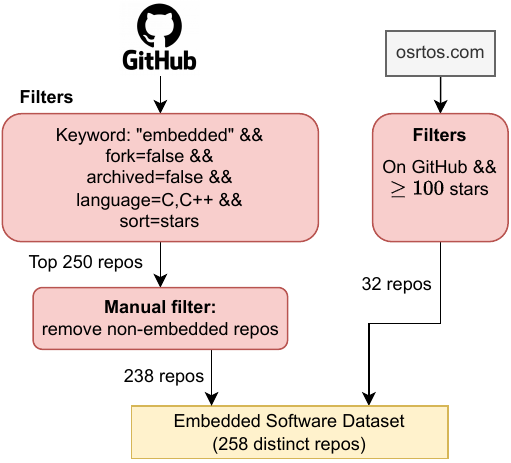}
    \captionof{figure}{
      Two-pronged approach to collecting embedded software dataset.
      The GitHub search (left side) was performed on April 8, 2023.
      The osrtos.com search (right side) was performed on June 7, 2023.
      }
    \label{fig:EMBOSSCorpus-Figure}
\end{wrapfigure}

We aim to collect a set of representative and well-engineered~\ac{EMBOSS}.
We combine two approaches (\cref{fig:EMBOSSCorpus-Figure}).
First, we searched GitHub for embedded software projects (\cref{sec:EMBOSSCorpus-Crawl}).
Second, we used an external index of RTOSes (\cref{sec:EMBOSSCorpus-RTOS}).

\subsubsection{EMBOSS from GitHub Search} \label{sec:EMBOSSCorpus-Crawl}
We searched GitHub for popular embedded software on GitHub.
Specifically, we collected original (\ie{} non-forked), active (\ie{} non-archived) C/C++ embedded software.
\cref{fig:EMBOSSCorpus-Figure} shows the exact filters for our search.
The initial query yields $\sim$20K projects.
We sorted them by popularity (operationalized as the number of stars~\cite{borges_whats_2018}) and collected the top 250.
We manually filtered out 12 false positives (non-embedded repositories) based on their READMEs.
For instance, we filtered out a machine learning project that had the word ``embedded'' in its keywords.

\subsubsection{EMBOSS from Index of RTOSes} \label{sec:EMBOSSCorpus-RTOS}

Embedded systems are usually powered by an~\ac{RTOS}, which provides the necessary library and scheduling support for various application components.
We collected~\acp{RTOS} from~\url{osrtos.com}~\cite{osrtosweb}, which maintains a curated list of open-source \acp{RTOS}.
Specifically, we selected those available on GitHub with >100 stars.
This resulted in a total of 32 repositories.

\subsection{Analysis of Corpus} \label{sec:EMBOSSCorpus-Analysis}

We combined the repositories and de-duplicated them, resulting in a total of \numrepos unique~\ac{EMBOSS} repositories.
\cref{tab:EMBOSSCorpus-Summary} summarizes all projects along with their fine-grained categorization (performed manually).
Most repositories are reasonably large, with a median of 33K~\ac{SLOC} and a maximum >400K~\ac{SLOC}.
This is similar to the project sizes examined in other studies~\cite{10.1145/3589610.3596271}.
\ifSUPPLEMENT
\fi

We also measured the repositories using the OSSF criticality measure (\cref{sec:Background-EMBOSS-OSSFCriticalityScore}).
All 13 categories have a median project with ``medium'' or ``high'' criticality score; the overall median criticality score is 0.43 (high).
This indicates that our corpus includes important projects.
\ifSUPPLEMENT
\fi

\subsection{Study Methodology} \label{sec:Theme1-RQ1-Methods}
We examine the state of practice usage of \ac{SAST} tools from two views:
  the use of SAST in the CI workflows of the corpus,
  and
  a survey of the project developers in the corpus.

\subsubsection{Workflow Analysis} \label{sec:Theme1-RQ1-Methods-WorkflowAnalysis}

We noticed that 42\% (109/\numrepos) of the \ac{EMBOSS} repositories use GitHub workflows to build and test the underlying codebase.
We automatically analyzed these workflows to detect the use of~\ac{SAST} tools.
Specifically, for each Action used in a workflow, we check if it is a~\ac{SAST} tool by checking its category in the GitHub CI Actions marketplace.
We define~\ac{SAST} tools as those whose marketplace category is ``code quality'' or ``security.''
Next, we manually check every matching Action to validate that it is indeed a~\ac{SAST} tool.

For workflows for which no SAST was found (248/258 of projects), we estimated whether or not this occurred due to errors in our automated analysis, or because they indeed used no SAST.
We performed a random sampling of~\numrandomsampling{} workflows and manually checked them.

To make the measured rate of SAST usage interpretable, we performed the same measurement on the top 5,000 OSS projects deemed ``critical'' and ``extremely critical'' according to the OpenSSF criticality score.
These projects do not target embedded contexts -- none of the projects from our corpus appear in this list.

\noindent\textbf{Results.} We found that only \numprojectsast (\numprojectsastperc) of the repositories use a sophisticated \ac{SAST} tool.
All of these use free \ac{SAST} tools, specifically, \codeql.
None of them use commercial \ac{SAST} tools.
Of the 10 repositories that use \codeql, 7 use an out-of-date version.

By comparison to the top 5,000 OSS projects by criticality (without the embedded constraint), we can see how small this adoption rate is.
Of the top 5,000 OSS projects we examined for comparison, ~\numtopprojsast{} (\numtopprojecsastperc{}) use $\geq$1~\ac{SAST} tools by our definition.

In our random sampling to check for false negatives in the \ac{EMBOSS} measure (a random sample of 20 projects), we found only two false negatives,~\ie{} 10\% false negative rate.
Both were due to a level of indirection around the use of a SAST tool.
\code{RIOT-OS/RIOT} runs its static tests in a Docker container, and~\code{InfiniTimeOrg/InfiniTime} runs \code{clang-tidy} in a script.

\subsubsection{Developer Survey} \label{sec:Theme1-RQ1-Methods-DeveloperSurvey}
\label{subsubsec:developersurvey}
To complement our previous workflow analysis, we conducted a developer survey under the supervision of our institution's Institutional Review Board (IRB). The survey aimed to gather insights from projects' maintainers about their security practices and to identify any alternative ways in which they might use~\ac{SAST} tools.
Our population of interest was the maintainers of the 248 (96\%) of projects that do not use any~\ac{SAST} Actions.
For each of these projects, we collected emails of users who recently contributed and emailed them the link to our survey. %
We were able to find the maintainers' email for~\numemailsent{} (out of 248) projects.
There were 15 questions in the survey with an anticipated time of 5 min.
The full survey is in~\apdx{apdx:survey-questions}.

\noindent\textbf{Results.} We got 25 responses (24\% response rate), representing 20 distinct repositories.
This response rate is comparable to that reported by other works that survey developers from GitHub (\eg~\cite{205176, jiang2017and}). While the survey's sample size is relatively small, it still offers valuable insights into developers' perspectives, and complements our quantitative measurement results for this research question.

\begin{figure*}[h!]
    \centering
    \includegraphics[width=0.9\textwidth]{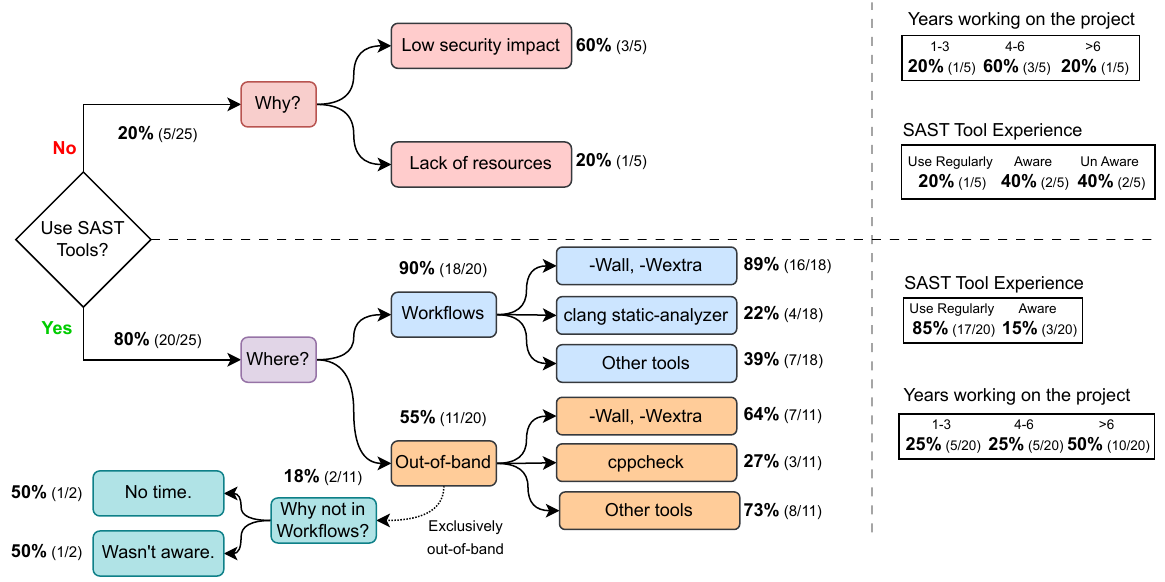}
    \caption{
    Summary of our developer survey on the use of~\ac{SAST} tools.
    }
    \label{fig:surveysummary}
\end{figure*}

\cref{fig:surveysummary} illustrates the responses to our survey.
We discuss why developers do and do not use SAST tools.

\myparagraph{Use of~\ac{SAST} Tools.}
Most of the surveyed developers (80\% (20/25)) claimed to be using~\ac{SAST} tools.
\emph{However, the most commonly used approach was enabling compiler warnings such as \code{gcc -Wall -Wextra}, and developers considered those to be adequate SAST.}
Compiler warnings are not effective as they mainly catch simple issues and have high false negative rates.
As a simple demonstration of the weaknesses of compiler warnings as SAST, we executed~\code{gcc}'s analyses on a set of test cases from the~\codeql{} repository.
These are simple test cases (<10 lines), each demonstrating a security issue, \eg~using~\simplecode{\%s} in~\code{scanf} or passing invalid pointer types to a function call.
We compiled these test cases using a recent version of~\code{gcc} (\latestgccversion{}) with strict warnings.
This configuration of \code{gcc} found issues in only~\numdefecttypesfound{} (\numdefecttypesfoundperc{}) defect types.
Some simple security issues were flagged, such as the use of~\code{strcpy} instead of~\code{strncpy}.
However, more complex ones were missed, such as inconsistent~\code{NULL} checks and use-after-free errors.
We provide more details in~\apdx{apdx:gcceffectiveness}.
This shows that current~\ac{SAST} practices in~\ac{EMBOSS} are not adequate.

\myparagraph{Not Using~\ac{SAST} Tools.}
20\% (5/25) of the developers use no~\ac{SAST} tools.
Most of these respondents (3/5) believe the security vulnerabilities in the corresponding projects have a low impact.
However, these projects have an average~\ac{OSSF} criticality score of 0.43 (``high'').
These developers may underestimate the severity of security issues in their projects, in line with previous studies~\cite{xie2011programmers, 8816991}.
The developers of another project reported insufficient resources (\eg time). 
Unfortunately, this project is one of the most popular (>5K stars) open-source C++ libraries for embedded systems, with an ~\ac{OSSF} score of >0.6 (``critical'').

A final common reason for non-SAST use was concern about their effectiveness.
Five respondents felt that using~\ac{SAST} tools on embedded software is questionable and might result in many false positives.
This finding is consistent with previous surveys of non-embedded software developers~\cite{johnsonWhyDonSoftware2013}.

To summarize our findings from this study of~\ac{SAST} prevalence in~\ac{EMBOSS}:
\begin{tcolorbox} [width=\linewidth, colback=yellow!30!white, top=1pt, bottom=1pt, left=2pt, right=2pt, float]
\textbf{Finding 1}: Sophisticated SAST tools are rarely used in \ac{EMBOSS} repositories.
According to our workflow analysis, only \numprojectsastperc of the \ac{EMBOSS} repositories do so.
With the same measure, \numtopprojecsastperc{} of non-embedded OSS do. %
Many~\ac{EMBOSS} repositories rely only on compiler warnings for SAST, which fail to find many common security defects.

\textbf{Finding 2}: The surveyed developers are generally aware of~\ac{CI} workflows and use them to run their~\ac{SAST} tools. When they do not use SAST, it is commonly because they believe the security impact or effectiveness of SAST is low.
\end{tcolorbox}

\section{SOTA SAST Performance}
Given the lack of \ac{SAST} tool usage in \ac{EMBOSS}, our aim is to understand the effectiveness of SOAT \ac{SAST} on \ac{EMBOSS}.

\subsection{Selection of the SOTA SAST Tools}
\label{subsec:selectionofsast}

{
\begin{table*}[ht]
\caption{
  Comparison of the SAST tools we considered, on the Juliet benchmark and our~\ac{EMBOSS} dataset.
  The median \# of warnings is reported for repositories where the tool ran successfully (\codeql did not produce warnings on over half of repositories).
  *\ac{EMBOSS} dataset precision is estimated via sampling.
  The cpp-lint performance measurement may be biased (see text).
  }
\label{tab:SASTToolComparison}
\centering
\scriptsize
\begin{tabular}{@{}ll|c|rlcc@{}}
\toprule
\textbf{GitHub Action} & \textbf{SAST}
  & \textbf{Juliet perf.}
  & \textbf{\# Repos} & \textbf{Failure(s)} & \textbf{Med. \# warn.} & \textbf{Precision} \\
\toprule
  david-a-wheeler/flawfinder & flawfinder~\cite{wheeler2006flawfinder}
    & Error
    & 176 (68\%) & Invalid SARIF; Crashes & 12 & 64/316 (20\%) \\
  cpp-linter/cpp-linter-action & cpplinter~\cite{lemos_cc_2023}
    & Timeout 
    & 230 (89\%) & Timeout; Crashes & 111 & 0/213 (0\%)* \\
  deep5050/cppcheck-action & cppcheck~\cite{pal_deep5050cppcheck-action_2023}
    & Timeout 
    & 256 (99\%) & Timeout & 19 & 116/200 (58\%) \\
  github/codeql-action & CodeQL~\cite{avgustinov2016ql}
    & F$_1$: 0.21 %
    & 74 (29\%) & Autobuild failure & 0 & 154/160 (\textbf{96\%}) \\
\bottomrule
\end{tabular}
\end{table*}
}

We want to apply the best-performing open-source SAST tool that (1) has a low false positive/negative rate;
(2) can be readily used on \ac{OSS} repositories;
(3) is stable (not pre-release), free (not requiring licenses), and
  ``plug-and-play'' (supports a range of compilers and does not require knowledge of program semantics/modeling, etc.).

\myparagraph{Competing Tools.}
We selected popular GitHub Actions that perform \ac{SAST} on C/C++ repositories, as shown in~\cref{tab:SASTToolComparison}.
Selection criteria are detailed in~\apdx{apdx:selection-of-sast-tools}.
We made a workflow for each Action to apply them uniformly to the benchmarks for comparison.

\myparagraph{Juliet Benchmark Performance.}
As one measure of effectiveness, we tested these tools on the Juliet Test Suite.
The Juliet Test Suite is a labeled dataset commonly used to test~\ac{SAST} tools~\cite{julietusedtoevalsasttools}.
It does not focus on embedded software, so this is a measure of performance on general C/C++ code that may not reflect performance on embedded code.
We used a time limit of 6 hours, the maximum time allowed for a job on many CI platforms, such as GitHub CI~\cite{githubusagelimits}.

The middle column of~\cref{tab:SASTToolComparison} shows the results.
Most tools either errored out or timed out. %
\codeql{} completed in 40 minutes.
\codeql{} raised 11,101 warnings with a precision of 71\% (7,904/11,101) and a recall of 12\% (7,904/65,263).\footnote{We count a reported flaw as a true positive if the reported location matches that of a ground truth bug.}

\myparagraph{EMBOSS Sample Performance.}
To obtain another vantage, we also ran the tools on the 258 repositories in our~\ac{EMBOSS} corpus.
We needed ground truth to evaluate the precision of each tool.
CodeQL produced 471 warnings, while the others produced between 4K-200K warnings.
It was infeasible to check them all.
Therefore, we randomly sampled warnings to check.
Specifically, for each tool, we randomly selected~\numrandomsamplingrepos{} repositories with $< $~\randomsamplingmaxnumwarnings{} warnings and manually checked each warning for those repositories.

The final columns of~\cref{tab:SASTToolComparison} show results.
\emph{\codeql{} has the highest precision by far, at 96\% -- unsurprising given its effectiveness on Juliet Test Suite.}
cppcheck and flawfinder had false positive rates $> 40$\%.
We recorded cpp-linter as having 100\% false positives.
This was likely a flaw in our sampling approach: all sampled warnings were related to compile-time issues that did not cause cpp-linter to error out, but we expect the sampling approach caused us to only examine warnings related to projects it struggled to compile.

Given that \codeql{} is the most effective tool, we next perform a thorough evaluation of \codeql{}'s effectiveness on \ac{EMBOSS}.
\subsection{Effectiveness of \codeql{} \label{sec:codeql-effectiveness}}
As shown in \tbl{tab:SASTToolComparison}, \codeql{} failed to run on  71\% of \ac{EMBOSS} repositories.
Specifically, the \texttt{Autobuild} phase of \codeql{} failed to handle the diverse build setup of these repositories.
We, therefore, manually created build scripts for all repositories based on their documentation and existing \ac{CI} workflows.

\myparagraph{Build Scripts Creation.}
 \label{meth:ApplyingCodeQL}
We made the build scripts cover as much of the codebase as possible (\eg{} by compiling all example applications and all supported architecture and boards whenever possible).
We successfully created build scripts for~\numreposbuilt{} (\numreposbuiltperc{}) repositories.
For the other~\notbuiltrepos{} repositories, the build instructions were either missing (17), too complex (\ie{} unavailable toolchains or dependencies) (49), or we could not get them to work (36).
This manual process took $\sim$45-60 minutes per repository.

\myparagraph{Analysis and Configuration Details of CodeQL.}
\codeql{} supports many suites (\ie{} collections of queries). %
There are three built-in suites for security scanning: \simplecode{default}, \code{cpp-security-extended}, and \code{cpp-security-and-quality}.
Each is a subset of the next, so we used the largest of these, \code{cpp-security- and-quality}, which contains 166 queries.
Despite the queries' effectiveness on non-embedded codebases,
Our preliminary analysis showed that (i) A few of these queries are not applicable to embedded software, and (ii) The risk of corresponding defects is low because of the lack of process support and OS abstractions.
We identified nine such queries and excluded them from our analysis. \tbl{tab:reason-for-ignoring} shows the list of queries and the corresponding rationale for their exclusion.

\begin{table}[ht]
\centering
\small
\caption{Reasons for excluding certain \codeql queries. ``Code readability'' means the query detects code readability issues but not defects.}
\label{tab:reason-for-ignoring}
\begin{tabular}{@{}ll@{}}
\toprule
\textbf{Query}                                   & \textbf{Reason for ignoring} \\ \midrule
cpp/path-injection                      & Inapplicable            \\
cpp/world-writable-file-creation        & Inapplicable        \\
cpp/poorly-documented-function          & Code readability         \\
cpp/potentially-dangerous-function\tablefootnote{The query ``cpp/potentially-dangerous-function'' checks for calls to gmtime, localtime, ctime and asctime. These functions are not thread-safe.}      & Low-risk (Lack of OS abstractions and arbitrary process support)           \\
cpp/use-of-goto                         & Code readability         \\
cpp/integer-multiplication-cast-to-long & Low-risk (Most embedded device configurations are 32-bit)            \\
cpp/comparison-with-wider-type          & Low-risk (Most embedded device configurations are 32-bit)          \\
cpp/leap-year/*                         & Low-risk            \\
cpp/ambiguously-signed-bit-field        & Low-risk            \\ \bottomrule
\end{tabular}
\end{table}

Furthermore, based on initial results, we modified three queries to improve their precision and ignore certain restrictions. 
First, we modified the~\code {cpp/stack-address-escape} query to ignore cases of assigning a function parameter of a pointer type to a non-local variable.
This usage is commonplace in practice and is unlikely to constitute a defect of significant concern as embedded systems usually have a fixed memory layout.
Second, we modified~\code{cpp/constant-comparison} to only report comparisons that are always false because we found that always-true comparisons are usually not defects in the~\ac{EMBOSS} context.
For instance, developers can be overly cautious and perform the same check multiple times, where the second check will always be true.~\eg{}~\code{... if (p != NULL) { ... if (p != NULL) ... }}.
Third, we modified~\code{cpp/uninitialized-local} to eliminate false positives caused by casting a variable explicitly to \code{void}.
Developers prevalently use such casts in \ac{EMBOSS} to suppress compiler warnings on unused variables, \eg~\code{(void) x;}.
\codeql{} accepted this third modification into their main repository~\cite{codeqlpullrequest}.
We did not propose the first two modifications to the \codeql{} team, as they may increase the false negative rate in general-purpose software analysis and are more appropriate as domain-specific refinements for embedded systems.

Finally, we created GitHub workflows for the \numreposbuilt{} successfully-built repositories.
These workflows invoke the necessary build scripts and run~\codeql{} with the required configuration.
We ran these GitHub workflows on these \numreposbuilt{} repositories.
This produced many~\textit{issues}, which~\codeql{} divides into
  \textit{errors} (high-severity concerns, \eg memory un-safety)
  and
  \textit{warnings} (lower-severity issues, \eg code smells).

\input{tables/mainresults}

\tbl{tab:codeql-results} shows the summary of our results across all repositories.
We discuss the results by answering the following research questions:
\begin{itemize}[topsep=0pt]
    \item[\textbf{RQ1}]{What defects does \codeql find in EMBOSS?}
    \item[\textbf{RQ2}]{How do results vary by~\ac{EMBOSS} type?}
    \item[\textbf{RQ3}]{What is the false positive rate of~\codeql?}
    \item[\textbf{RQ4}]{How do developers respond to~\codeql results?}
    \item[\textbf{RQ5}]{Will developers integrate~\codeql in CI pipelines?}
\end{itemize}

\subsubsection{RQ1: Defects Identified by \codeql in EMBOSS}

\input{fig/cdfbugsperrepo}
We manually analyzed all~\codeql{} issues for~\numreposanalyzed{} repositories (out of a possible~\numreposbuilt{}).
The others have a substantial number of issues, and we did not have time to analyze them thoroughly.
As reported in the~\textit{Defects Discovered} row of~\cref{tab:codeql-results}, we identified \numalldefects defects across \numreposwithdefect repositories.
We also distinguish the proportion of defects that can be deemed security-relevant to understand whether the studied~\codeql{} query suite has security benefits (\eg{} memory safety issues) vs. broader quality benefits (\eg{} code smells).
To be conservative, we define a defect as \textbf{security-relevant} if and only if (1) the \codeql query finding the defect contains the security tag, or (2) the defect is clearly related to memory safety (\eg null pointer dereference, out-of-bounds read/write).
There were~\numsecuritydefects{} (\numreposwithsecdefect{} repositories) security-relevant defects, including in major projects maintained by organizations like Microsoft, Amazon, and the Apache Foundation.
\ac{EMBOSS} engineers have confirmed~\numallconfirmeddefects{} (\numallconfirmeddefectsperc{}) of these defects, mainly by accepting our pull requests.

\myparagraph{Defect Rates Per Repository.}
\fig{fig:cdf-bugs-per-repo} shows \acp{CCDF}  of the number of total defects and security defects in each repository.\footnote{A point $(x,y)$ on a \ac{CCDF}~\cite{wikipedia_contributors_cumulative_2025} line indicates that $y$\% of repositories contain greater than $x$ corresponding type of defects.}
The left-most point on both the lines indicates that there are~\percentreposwithdefect{} (\numreposwithdefect{}) repositories with at least one defect, and~\percentreposwithsecdefect{} (\numreposwithsecdefect{}) repositories with at least one security defect.
The security defects line has almost the same trend as total defects, which is consistent with our finding that a large proportion (\numsecuritydefects{} out of \numalldefects{}, or \numsecuritydefectsperc{}) of the identified defects are security-relevant.
Although \textasciitilde90\% of the repositories have less than ten total defects, nine repositories have significantly more.
\tbl{tab:topreposwithdefects} lists the top 5 repositories with the most total defects and their criticality scores.

\input{tables/topreposwithdefects}

\myparagraph{Common Types of Security Defects.}
We found several classes of security defects across all repositories.
\fig{fig:bugs-per-rule} shows the top 10 types of security-relevant defects~\cite{github_inc_codeql_cpp_doc_2023} found along with the corresponding number of defects.
We discuss three main types of security defects below. Code examples for each type are provided in~\apdx{apdx:commontypesofsecuritydefects}.

\input{fig/codeqlqueryfig}

\fakeitem \textit{cpp/inconsistent-null-check}. This rule identifies cases where the return value of a function is not checked for~\code{NULL}, despite most other calls to the same function performing such a check. Developers should consistently validate return values that may be~\code{NULL} to prevent potential null pointer dereferences. This rule flagged 135 such instances.
This rule detected 135 such instances.

\fakeitem \textit{cpp/uncontrolled-allocation-size}.
This rule detects cases where the size argument of a memory allocation function (\eg~\code{malloc}) is computed through integer arithmetic involving potentially untrusted input (\eg~user input). If the input takes on large values, an integer overflow~\cite{dietz2015understanding} may occur, resulting in an allocation size significantly smaller than intended. Subsequent buffer accesses may lead to out-of-bounds reads or writes. This rule identified 49 such instances.

\fakeitem \textit{cpp/unbounded-write}. This rule detects out-of-bound write vulnerabilities.
Specifically, this includes analysis of potentially dangerous function calls (\eg{}~\code{strcpy}, \code{sscanf}) to check whether these are used properly with valid arguments.
This rule detected 47 vulnerabilities of potential buffer overflow.

\input{fig/cdfbugseverity}
\myparagraph{Severity of Security Defects.}
The severity of a security bug depends on its exploitability and the criticality of the underlying software~\cite{10.1145/2989238.2989239, 9825784}.
Given the large number of defects, manually assessing exploitability is intractable.
Instead, we use the OSSF criticality score (\sect{sec:Background-EMBOSS-OSSFCriticalityScore}) of the target repository to assess the severity of a bug.
\fig{fig:cdf-bug-severity} shows the \ac{CCDF} of the severity of security defects.
Specifically, a point $(x,y)$ on the line indicates $y$\% of the defects have severity greater than $x$.
Approximately 50\% of bugs have a severity score of more than 0.5, which represents high-severity repositories (\sect{sec:Background-EMBOSS-OSSFCriticalityScore}).
Specifically, \textasciitilde40\% of bugs have a score of more than 0.6, representing vulnerabilities in critical repositories.
For instance, we found an arbitrary write vulnerability in \code{Mbed-TLS/mbedtls} (criticality score = 0.73, \lst{lst:unbounded-write-mbedtls}) and a use-after-free in \code{apache/nuttx} (criticality score = 0.69, \lst{lst:useafterfreeinsensor}); both of these are critical projects.

\vspace{0.5cm}
\input{listings/unboundedwrite}\hfill
\input{listings/useafterfreelisting}
\vspace{0.5cm}

\input{fig/nonsecbugsperrule}

\myparagraph{Common Types of Non-Security Defects.}
These defects may not lead to security vulnerabilities but can cause functionality issues, undefined behavior, and compilation issues.
For instance, the rule \simplecode{cpp/missing-return} detects non-void functions with no explicit return statement.
This may result in undefined behavior during runtime~\cite{linearity_omitting_2010}.
Similarly, the rule \simplecode{cpp/virtual-call-in-constructor} detects calls to virtual functions in a constructor.
This also could lead to undefined behavior as the object's virtual table may not be completely initialized~\cite{github_inc_virtual_2023}.
\fig{fig:non-sec-bugs-per-rule} shows the top ten non-security defects along with the corresponding number of defects.

\subsubsection{RQ2: Trends by~\ac{EMBOSS} Type}
\cref{fig:vultypesinemboss} shows the number of defects found across various repositories according to their categories.
\emph{At a high level, across all categories, the number of security defects is more than that of the number of non-security defects.}
Furthermore, the number of defects is proportional to the number of repositories of the particular category (\cref{tab:EMBOSSCorpus-Summary}).
For instance, Network (NET), Operating Systems (OS), and Applications (APP) are the top three categories containing the highest number of repositories (128 (50\%)), and they also contain the highest number of defects (423 (60\%)).
The Memory Management Libraries (MML) with the least number (4) of repositories also have the least defects (6).
Interestingly, we noticed that defect density,~\ie{} number of defects per KSLOC, is non-uniform.
\apdx{apdx:defect_density} provides defect distribution per-repository and defect density across various categories of~\ac{EMBOSS}.
In summary, APP and NET have the highest defect densities. On the other hand, OS and HAL have the lowest densities.
\emph{Our results empirically show defect density is not uniform across different categories of~\ac{EMBOSS}}.

\input{fig/bugssplit}

\subsubsection{RQ3: False Positive Rates}
False positive rate analysis requires a significant amount of work, yet false positives are also a major concern in the adoption of SAST tools.
Given the large number of repositories, we sampled \numreposfortpfp successfully built repositories (the 50 most starred, the 50 least starred, and 23 randomly picked repositories) and manually categorized all issues in them into true and false positives.
A false positive means that the result does not match what the rule intends to detect, \eg an error for an uninitialized variable when it is actually initialized.
Two analysts worked for one month to analyze the results.
The two analysts worked largely independently but discussed uncertainties with each other and with the rest of the research team.
All analysts and researchers had substantial training (coursework and experience) in C/C++ programming and cybersecurity, and we believe they were able to make correct judgments about whether or not a~\codeql{} issue represented a true positive.

The overall percentages of true and false positives are \codeqltpperc~(\codeqlnumtp/\codeqlnumwarnsampledfortpfp) and \codeqlfpperc~(\codeqlnumfp/\codeqlnumwarnsampledfortpfp), respectively.
\fig{fig:cdf-tp-fp} shows the \ac{CDF} of the false positive rates of different rules.
Specifically, a point $(x,y)$ on a line indicates $y$\% of the rules have false positive rates of less than or equal to $x$\%.
Approximately 60\% of the rules had no false positives, and 10\% had no true positives.
This indicates that false positives are polarized, and a few rules contribute to the majority of false positives.
Specifically, 20\% of rules contribute to more than 60\% of false positives.
We discuss below the top four~\codeql{} queries contributing to false positives. Comprehensive information on all rules contributing to false positives is provided in~\apdx{apdx:commonfprules}.

\begin{itemize}[leftmargin=*]
    \item \textit{cpp/uninitialized-local}. Dataflow analysis of \codeql{} is not path-sensitive. Some variables may not be initialized in all paths. However, when a variable is used, certain path conditions hold, under which it can be proved that the variable must have been initialized. %
    \item \textit{cpp/missing-check-scanf}. Developers can use \code{switch case} statements (instead of \code{if} statements) to check the return value of \code{scanf} calls. These are valid checks but not detected by \codeql{}.
    \item \textit{cpp/suspicious-pointer-scaling} and \textit{cpp/suspicious-pointer-scaling-void}. These rules detect risky pointer arithmetic operations. However, pointer casts, and type-punning are pretty common and unavoidable in low-level embedded system code.
    \item \textit{cpp/unbounded-write}. \code{strcpy} is safe if the destination must be large enough. For example, developers can first use \code{strlen} to calculate the length of the source string, allocate enough memory for the destination string, and then call \code{strcpy}.
\end{itemize}

Although the cumulative false positive rate is high (\codeqlfpperc), it does not affect most repositories.
\fig{fig:fp-perc-per-repo-cdf} shows the \ac{CDF} of \% of repositories and false positive rate; we can see that $\sim$40\% of repositories have no false positives and more than 60\% of the repositories have less than 20\% false positive rate.
Furthermore,~\emph{the actual number of false positives is very low, as shown in~\fig{fig:fp-num-per-repo-cdf}. Specifically, $\sim$55\% of the repositories have less than one false positive, and 90\% of repositories have less than ten false positives}.
These results show that the majority of~\ac{EMBOSS} repositories are not affected by false positives.
\input{fig/falsepositivesperrulesfig}

\subsubsection{RQ4: Developer Response on Defects Identified by \codeql{}}
\label{sec:developer-response}
We responsibly disclosed all identified defects in repositories that are actively maintained (had commits in the past three months).
We opened issues and raised pull requests with appropriate patches where possible.
The bottom of~\tbl{tab:codeql-results} summarizes the developer response. %
In total,~\percentallconfirmeddefects (\numallconfirmeddefects/\numalldefects) of defects have been confirmed by developers (via merging our pull requests or expressing confirmation in replies to issues).

Most of the patches were readily accepted by the developers.
In a few cases, developers were even interested in knowing the techniques we used to find the defects.
For instance, developers of an AWS-owned repository
said~\emph{``I'm curious how you stumbled across this -- Was there some sort of test you ran or was this something that came up during your development? I'm hoping we can duplicate your method of discovery to add some sort of check/test to the repo.''}

There were two pull requests where the developers did not choose to fix potential security issues.
They stated that although code robustness is important, they deemed reduced code size and RAM usage to be a higher priority in their embedded software.
These observations support the conventional wisdom that software engineers (and especially engineers in embedded systems) trade-off between security and performance~\cite{fujdiak_seeking_2018,gopalakrishnaIfSecurityRequired2022}.

Although many security-relevant defects were resolved, only two ~\acp{CVE} were assigned.
When we disclosed the security-relevant defects, we did not explicitly ask the engineering teams to issue~\acp{CVE}.
Of the 94 repositories against which we opened at least one security-relevant defect, only two issued~\acp{CVE} for these defects:
  \code{mbedtls} issued CVE-2023-29472,
  and
  \code{contiki-ng} issued CVE-2023-30546.
We eventually followed up on our 77 reports of defects to the 10 most popular repositories (by GitHub stars) to inquire whether~\acp{CVE} were being prepared.
Two of the engineering teams replied suggesting that we email their security teams -- we did so, but received no response.
The other eight teams did not respond.
Our research supports the observation of prior work~\cite{9152613}, that security defects are often fixed ``silently'', without tracking via a~\ac{CVE}.\footnote{With the recent threat to funding for the CVE system~\cite{vicens_us_2025,satter_last-minute_2025},
this finding provides further evidence that CVEs may not be the best metric for assessing the impact of a work.}

\subsubsection{RQ5: Developer Response on Integrating \codeql Workflows} \label{sec:rq-codeql-wf}
We opened pull requests to integrate our~\codeql{} scanning workflows into the CI pipeline of the projects.
This would have the effect of using~\codeql{}'s~\ac{SAST} to check all subsequent pull requests.
We measured the number of merged pull requests and the kinds of replies made by the developers.
An example pull request is given in~\apdx{apdx:ExamplePR}.

We raised \numcodeqlworkflowpullquests pull requests to integrate our~\codeql{} workflows into the corresponding projects.
We did not submit some pull requests as the repositories do not accept external contributions,~\eg{} Microsoft Azure.
In addition, some of our workflows became out of date due to concurrent changes in the project's build process. 
\apdx{wfpullrequestsummary} shows our pull request.
We received responses for 52 of our pull requests, of which~\nummergedworkflows were merged (71\% acceptance rate for responses, \nummergedworkflowsperc acceptance rate overall).

\myparagraph{Accepted Requests.}
Most of the developers readily accepted our workflow.
In a few cases (3), we had to make syntactic adjustments %
to our workflow according to the repository coding practices.
Few developers (2) had concerns of the effectiveness of~\codeql{}.
When asked, we pointed to the defects we identified as evidence. %
Interestingly, one developer surveyed their friends on X (formerly Twitter) for opinions about~\codeql{}, before accepting our pull request.

\myparagraph{Closed Requests.}
Several developers (7) closed our pull requests, assuming that these were generated by bots.
We contacted them again to clarify that we were not bots but received no response.
A few developers (3) mentioned that they do not have enough resources to handle the alerts raised by~\codeql{}.
A few developers (2) mentioned concerns about licensing.

In summary, this part of our investigation yielded the following findings:

\begin{tcolorbox} [width=\linewidth, colback=yellow!30!white, top=1pt, bottom=1pt, left=2pt, right=2pt]

\textbf{Finding 3 (RQ1)}: \codeql{} finds hundreds of real defects in the studied~\ac{EMBOSS} repositories, including in repositories maintained by reputable organizations like Amazon and Microsoft.

\textbf{Finding 4 (RQ2)}: Defect density (defects per SLOC) is not uniform across different categories of~\ac{EMBOSS}. Some categories of projects (\eg{} APP and NET) are more likely to contain defects than others (\eg{} OS and HAL).

\textbf{Finding 5 (RQ3)}: \codeql{} has a false positive rate of \codeqlfpperc in the \numreposfortpfp sampled repositories. 
However, false positives are polarized,~\ie{} a few rules contribute to the majority of false positives.

\textbf{Finding 6 (RQ3)}: Although the overall false positive rate is high, it has minimal impact on~\ac{EMBOSS} repositories: $\sim$40\% of repositories have no false positives, $\sim$55\% of the repositories have $\leq1$ false positive, and 90\% of repositories have $\leq10$ false positives.

\textbf{Finding 7 (RQ4)}: Developers readily accept fixes for defects identified by \codeql{} -- demonstrating that they care about these defects.

\textbf{Finding 8 (RQ5)}: Many~\ac{EMBOSS} developers are willing to integrate the CodeQL SAST into their projects' CI as a GitHub workflow, provided that someone else (our research team) prepares, validates and explains the workflow for them.

\textbf{Finding 9}: A default Autobuild fails on many~\ac{EMBOSS} projects.
However, producing a customized build suitable for~\codeql{} takes minimal engineering effort for developers.
\end{tcolorbox}

\section{Lessons Learned}
\label{sec:lessonslearned}

We summarize our experiences in \textbf{four lessons} on using~\ac{SAST} in~\ac{EMBOSS}.

\myparagraph{(Lesson 1) \ac{EMBOSS} can benefit from \ac{SAST}:} 
Despite developers' misgivings about the effectiveness of \ac{SAST} on \ac{EMBOSS}, we found many security defects (\numsecuritydefects{}) across various embedded software by using an existing~\ac{SAST} tool.
Developers acknowledged and fixed most of the security defects (\percsecdefectsconfirmed) found by~\ac{SAST} tools, which shows that~\ac{SAST} tools can find important defects.
Since many of these repositories (\numprojectnosastperc) did not use~\ac{SAST} tools, it is perhaps unsurprising that they were rife with defects that~\ac{SAST} can detect.
Nevertheless, \textit{evidence} of this is important to push the~\ac{EMBOSS} engineering community toward more responsible engineering practice.

\myparagraph{(Lesson 2) Developers are willing to adopting \ac{SAST} in \ac{EMBOSS} Repositories:}
Several developers accepted our pull requests (\percmergedwfactiverepos{} (Active) and \nummergedworkflowsperc{} (Total)) to integrate a \ac{SAST} tool (\ie{}~\codeql{}) into their CI pipeline.
Our pull request was well-formatted and included all the necessary details along with evidence of \codeql{}'s effectiveness.
Specifically, we included the examples of the defects found by \codeql{} in the corresponding repository.
Furthermore, there was not much persuasion needed to accept our pull requests.
We draw two sub-lessons here.
First, engineers can easily integrate \ac{SAST} tool into \ac{EMBOSS} repositories -- the pull request is not too complex and can be done without much project-specific expertise.
Second, engineers will accept contributions from researchers, provided the contributions come with a demonstration of effectiveness (\ie an acceptable cost-benefit tradeoff).

\myparagraph{(Lesson 3) \ac{SAST} tool developers should consider the properties of~\ac{EMBOSS}:}
Our experience shows that certain \ac{SAST} queries, which are effective on traditional (non-embedded) codebases, might be ineffective or inapplicable for embedded codebases.
We therefore recommend that \ac{SAST} tool developers take the characteristics of embedded codebases into consideration while evaluating their tool design decisions.
Part of our contribution is a set of modifications and configurations of \codeql{} queries that demonstrate the kinds of changes that are needed. %

\myparagraph{(Lesson 4) We need more best-effort defect detection techniques for \ac{EMBOSS}:}
We were able to find a large number of defects (\numalldefects{}), including security vulnerabilities (\numsecuritydefects{}), in \ac{EMBOSS} repositories by just using an off-the-shelf \ac{SAST} tool.
Our results complement a recent work~\cite{emnesttest} that used simple systematic testing to find several severe security issues in popular~\ac{EMBOSS} network stacks.
These works provide strong evidence that the \ac{EMBOSS} engineering community should investigate the potential of integrating simple or best-effort defect detection techniques.

\section{Future Work} \label{sec:Discussion}

Developers accepted our pull requests to integrate a \ac{SAST} tool (\ie{}~\codeql{}) into their CI pipeline.
However, this required manual effort (although minimal) to identify the build setup, create a CI workflow, and raise the pull request.
As part of our future work, we plan to automate this process by using \acp{LLM} assisted techniques~\cite{10.1145/3664476.3664497}.
To further encourage the adoption of \ac{SAST} tools, we plan to create 
rewards badges (such as the OpenSSF Best Practices Badge~\cite{the_linux_foundation_badgeapp_2024}), or public recognition for projects that demonstrate the successful use of \ac{SAST} tools in finding and fixing vulnerabilities.
We also plan to create tutorials, workshops, and documentation that showcase the effectiveness of \ac{SAST} tools in identifying real-world vulnerabilities that can help \ac{EMBOSS} developers better understand their value.

\section{Limitations and Threats to Validity}

Like any empirical study, our study has a range of limitations.
We distinguish three types of threats to validity~\cite{wohlin2012experimentation}.
Guided by Verdecchia \etal, we focus on substantive threats that might influence our findings~\cite{verdecchia2023threats}.

\textbf{Construct Threats} are potential limitations of how we operationalized concepts.
We scope the construct of security vulnerabilities to those detectable by the~\ac{SAST} tools from the GitHub Marketplace, particularly those captured by \codeql{} queries with the security tag or those associated with memory safety.
Other classes of security vulnerabilities, and other kinds of software defects, are beyond the scope of our work.

\textbf{Internal threats} are those that affect cause-effect relationships.
This work was primarily a measurement study, which does not involve causal inferences.
However, our motivation stemmed in part from the observation that many~\ac{EMBOSS} projects do not use~\ac{SAST}, and that the surveyed developers often cited the perceived complexity and noisiness of applying~\ac{SAST}.
Our measurements are thus useful in shaping software engineering practice only insofar as these statements are truthful.

\textbf{External threats} may impact generalizability.
Here is where most of the threats are.
\begin{itemize}
\item \textit{Focus on free SAST tools:} We applied~\ac{SAST} tools available in the GitHub Marketplace to the open-source embedded software available on GitHub.
Our results may not generalize to other SAST tools, particularly commercial ones such as Coverity and Sonar.
\item \textit{Focus on~\ac{EMBOSS}:} Our results may not generalize to other embedded software, particularly commercial embedded software, to which costly techniques such as formal methods may have been applied~\cite{amusuo2025unit,amusuo2025unitTechReport}.
To shed some light on this threat, in our analysis, we showed that a \ac{SAST} tool (\ie{}~\codeql{}) was still able to find defects in commercially-developed open-source software, such as Amazon's \code{aws/aws-iot-device-sdk-embedded-C} (which uses the commercial Coverity SAST tool).
\item \textit{Scoping to GitHub:} Our study may suffer from data collection bias as we focus on projects and~\ac{SAST} tools available on GitHub.
There could be other~\ac{EMBOSS} projects (\eg{} in BitBucket) and tools on which our observations may not hold.
We tried to avoid this by collecting diverse projects with varying sizes.
\item \textit{Limited developer study:} Given the low number of responses, the observations from our developer study (\sect{subsubsec:developersurvey}) may not generalize to other~\ac{EMBOSS} repositories.
As a modest mitigation, we note that the response rate was consistent with other surveys of GitHub developers~\cite{205176, jiang2017and}.
\end{itemize}

\section{Related Work}
Earlier we discussed directly related work.
Here we compare broadly.

\myparagraph{Embedded Operating Systems and Frameworks:}
Al-Boghdady \etal~\cite{s21072329} conducted a thorough analysis of four IoT Operating Systems, namely
  RIOT, %
  Contiki, %
  FreeRTOS, %
  and
  Amazon FreeRTOS. %
  Their results indicated an increasing trend in the number of security errors over time.
  Others agreed:
    Alnaeli~\etal~\cite{alnaeli_vulnerable_2016, alnaeli_source_2017} reported a rise in unsafe statements in Contiki and TinyOS,
    and
    McBride~\etal~\cite{10.5555/3234847.3234913} found increasing error rates in Contiki. %
  Malik~\etal~\cite{malik_empirical_2023} shed some light on root causes, noting that the complex behaviors of embedded devices are challenging to validate internally. %
Our work encompasses these OSes and includes a wider range of embedded software, leading to a broader view of the state of~\ac{EMBOSS}.

\myparagraph{Other Analyses of Embedded Systems:}
Embedded software has been studied for decades.
We highlight a few recent analyses.
Peng~\etal~\cite{peng2024application} proposed a \ac{CI} environment to improve the efficiency and quality of software development in the nuclear power industry.
The XANDAR project~\cite{10546852} combines a model-based toolchain and hypervisor-based runtime architecture to create embedded software systems with safety, security, and real-time properties.
Bagheri~\etal~\cite{bagheri_synthesis_2020} proposed a method for automatically generating assurance cases for software certification.
Jia~\etal~\cite{contexiot17} used control and data flow analysis to find malicious behavior in IoT applications. %
Celik~\etals SOTERIA system~\cite{215955} combines static analysis and model checking to find security and safety violations in IoT software. %
Complementing these studies, we focused on static analysis for embedded software to understand current practices, challenges, and opportunities.

\myparagraph{Developers' Perspectives on~\ac{SAST} Tools:}
For SAST, many works have examined the factors hindering or spurring adoption.
Johnson~\etal~\cite{johnsonWhyDonSoftware2013}
found that false positives and (non-)usability of warnings are barriers.
More recently, Ami~\etal~\cite{ampn24} interviewed 20 practitioners and found that 
they considered these tools to be highly beneficial complements to manual analysis. %
Among the challenges faced by developers, the significant pain points were false negatives, the absence of meaningful alert messages, and the effort required for configuration and integration.
Wadhams~\etal~\cite{wadhams2024barriers} identified false positives, poor output, time-consuming setup, and manual effort for fixes as the primary barriers to \ac{SAST} adoption. They emphasized that both developers and \ac{SAST} tool creators have distinct yet equally crucial roles in promoting the widespread use of \ac{SAST}.
Our study revealed slightly different findings.
In addition to false positives, developers were unaware of the effectiveness of~\ac{SAST} tools on embedded software.

In terms of tool performance, 
Lenarduzzi~\etal have questioned the tools' capabilities~\cite{LENARDUZZI2023111575}, comparing six~\ac{SAST} tools for Java and finding little agreement among them as well as low precision.
Our experience contrasts with their findings.
Our experiments with \codeql{} demonstrate that SAST tools are highly capable of identifying vulnerabilities within the \ac{EMBOSS} context.
We were able to easily (with minimal engineering effort) configure and integrate~\codeql{} in~\ac{EMBOSS} repositories.
The alert messages were displayed in SARIF format and were easy to understand and evaluate.

\section{Conclusions}

We evaluated the usage and effectiveness of \ac{SAST} in \ac{EMBOSS}.
Across~\numrepos{} open-source embedded software projects, the \codeql{} \ac{SAST} tool found \numalldefects defects (with a false positive rate of \codeqlfpperc), \numallconfirmeddefects of which have been confirmed.
These included \numconfirmedsecdefects defects that were security vulnerabilities such as crashes and memory corruption.
False positives were mainly caused by a few outlier \codeql rules and projects. For the majority of repositories studied, the false positive rates were low.
We also raised pull requests to incorporate our \codeql{} workflows as part of \ac{EMBOSS} \ac{CI} pipeline, out of which~\nummergedworkflows{} (\percmergedwfactiverepos{} (Active) and \nummergedworkflowsperc{} (Total)) are already accepted.
We conclude that the current generation of static analysis tools, exemplified by~\codeql{}, has overcome concerns about false positives and can be easily incorporated into embedded software projects.
If engineers adopted these tools, many security vulnerabilities would be prevented.
\textit{Future research should push the bounds of vulnerability discovery, but we call for efforts to promote adoption of existing tools.}

\section*{Research Ethics} \label{sec:ethics}

In the conduct of this study, we upheld two ethical duties:
  the responsible conduct of research on human subjects,
  and
  the appropriate handling of cybersecurity vulnerabilities.

\myparagraph{Ethics for human-subjects research:}
Studies of human subjects must offer a favorable risk-reward tradeoff.
Our study included a human-subjects study: we surveyed~\ac{EMBOSS} software engineers.
The possible risk to our subjects was professional scrutiny based on following (or not following) best practices in software engineering such as using SAST.
The benefit is an increased awareness of the available SAST tools and their performance, which may benefit them directly, as well as those who depend on their software, and the broader~\ac{EMBOSS} community.

This study was conducted with the approval of our institution's Institutional Review Board (IRB).

\myparagraph{Ethics for cybersecurity vulnerabilities:}
The ethical duty for handling cybersecurity vulnerabilities requires responsible disclosure to protect users and systems by informing relevant parties about identified security risks in a timely manner.
Responsible disclosure typically follows one of two models: Coordinated Vulnerability Disclosure (CVD) or Full Disclosure. CVD involves informing the responsible parties first, allowing them time to address the issue before publicizing the vulnerability. Full Disclosure, in contrast, is the practice of publishing vulnerability analyses as soon as possible, without a private coordination period with the affected project or organization.

Since we identified vulnerabilities without associated exploits, and these vulnerabilities could be found by anyone applying a \ac{SOTA} tool, we determined that secrecy was not necessary. We therefore adopted the Full Disclosure approach by opening public issues or pull requests with patches for the identified vulnerabilities, as detailed in~\cref{sec:developer-response}.
By supplying a patch alongside the vulnerability report, we actively mitigated the risk to users by making it easier for maintainers to address the issue promptly. \ac{EMBOSS} engineers frequently fixed the vulnerabilities we identified, and none raised concerns that our public reporting was unethical.

\section*{Data Availability} \label{sec:dataavilability}

All project data are available at \url{https://github.com/purs3lab/ISSTA-2025-EMBOSS-Artifact} and \cite{shen_2025_15200316}.

\section*{Acknowledgments} \label{sec:acknowledgments}
This research was supported by the National Science Foundation (NSF) under Grants CNS-2340548 and CNS-2247686 and Rolls-Royce Grant on ``Dynamic Analysis of Embedded Systems.'' The U.S. government is authorized to reproduce and distribute reprints for Governmental purposes, notwithstanding any copyright notation thereon. Any opinions, findings, conclusions, or recommendations expressed in this material are those of the author(s) and do not necessarily reflect the views of the NSF or Rolls-Royce.

\clearpage

\bibliographystyle{ACM-Reference-Format}
\bibliography{bib/final-refs}

\ifSUPPLEMENT
\appendix
\input{newappendix-issta2025}
\else
\fi

\end{document}

%% file: macros.tex
\ifDEBUG
\newcommand{\machiry}[1]{\textcolor{green}{\textbf{Machiry:} #1}}
\newcommand{\JD}[1]{\textcolor{blue}{\textbf{Davis:} #1}}
\newcommand{\BY}[1]{\textcolor{brown}{\textbf{Yuan:} #1}}
\newcommand{\MS}[1]{\textcolor{violet}{\textbf{Ming:} #1}}
\newcommand{\AP}[1]{\textcolor{olive}{\textbf{Akul:} #1}}
\else
\newcommand{\machiry}[1]{}
\newcommand{\JD}[1]{}
\newcommand{\BY}[1]{}
\newcommand{\MS}[1]{}
\newcommand{\AP}[1]{}
\fi

\acrodef{MCU}{Micro Controller Unit}
\acrodef{HAL}{Hardware Abstraction Layer}
\acrodef{MMIO}{Memory Mapped Input Output}
\acrodef{KSLOC}{Kilo Source Lines of Code}
\acrodef{RTOS}{Real Time Operating System}
\acrodef{NCMA}{Non Conventional MMIO Access}
\acrodef{CFG}{Control Flow Grpah}
\acrodef{IR}{Intermediate Representation}
\acrodefplural{NCMA}[NCMAs]{Non Conventional MMIO Accesses}
\acrodefplural{RTOS}[RTOSes]{Real Time Operating Systems}
\acrodef{SAST}{Static Application Security Testing}
\acrodef{CI}{Continuous Integration}
\acrodef{EmS}{Embedded Software}
\acrodef{OSS}{Open-Source Software}
\acrodef{EMBOSS}{Open-Source Embedded Software}
\acrodef{SoTP}{State of The Practice}
\acrodef{SOTA}{state-of-the-art}
\acrodef{CWE}{Common Weakness Enumeration}
\acrodef{SLOC}{Source Lines of Code}
\acrodef{CVE}{Common Vulnerabilities and Exposures}
\acrodefplural{CVE}[CVEs]{Common Vulnerabilities and Exposures}
\acrodef{OSSF}{Open Source Security Foundation}
\acrodef{SARIF}{Static Analysis Results Interchange Format}
\acrodef{LLM}{Large Language Model}
\acrodef{CDF}{Cumulative Distribution Function}
\acrodef{CCDF}{Complementary Cumulative Distribution Function}

\newcommand{\codeql}{{\sc CodeQL}\xspace}

\newcommand{\code}[1]{%
  \mintinline[fontsize=\footnotesize{},mathescape, escapeinside=||]{cpp}{#1}%
}

\newcommand\encircle[1]{\tikz[baseline=(X.base)] \node (X) [draw, shape=circle, inner sep=0em,text width=1em, text centered] {\scriptsize #1};}

\newcommand{\tbl}[1]{Table~\ref{#1}}
\newcommand{\sect}[1]{\S\ref{#1}}
\newcommand{\fig}[1]{Figure~\ref{#1}}
\newcommand{\lst}[1]{Listing~\ref{#1}}

\ifSUPPLEMENT
 \newcommand{\apdx}[1]{Appendix~\ref{#1}}
\else
 \newcommand{\extendedreport}{Our Extended Report~\cite{extendedreport}\xspace}
 \newcommand{\apdx}[1]{\extendedreport}
\fi

\newcommand{\etal}{\textit{et al.}\xspace}
\newcommand{\etals}{\textit{et al.}'s\xspace}
\newcommand{\ie}{\textit{i.e.,}\xspace}
\newcommand{\eg}{\textit{e.g.,}\xspace}
\newcommand{\latestgccversion}{11.4.0\xspace}

\newcommand{\numdefecttypes}{82\xspace}

\newcommand{\numdefecttypesfound}{17\xspace}
\newcommand{\numdefecttypesfoundperc}{21\%\xspace}

\newcommand{\simplecode}[1]{{\footnotesize\texttt{#1}}\xspace}

\newcommand{\myparagraph}[1]{\vspace{0.15cm}\textbf{#1} \noindent{}}

\newcommand{\fakeitem}{\noindent\textbullet\xspace}

\usepackage{tcolorbox}

%% file: data/data.tex
\newcommand{\numsast}{12\xspace}
\newcommand{\numrepos}{258\xspace} %
\newcommand{\numreposbuilt}{156\xspace} %
\newcommand{\numreposbuiltperc}{60\%\xspace} %
\newcommand{\numreposanalyzed}{151\xspace} %
\newcommand{\notbuiltrepos}{102\xspace} %
\newcommand{\numalldefects}{709\xspace}
\newcommand{\numsecuritydefects}{535\xspace}
\newcommand{\numsecuritydefectsperc}{75\%\xspace}
\newcommand{\numallconfirmeddefects}{376\xspace}
\newcommand{\numallconfirmeddefectsperc}{\percentallconfirmeddefects}

\newcommand{\numprojectsast}{10\xspace}
\newcommand{\numprojectsastperc}{4\%\xspace} %
\newcommand{\numprojectnosastperc}{96\%\xspace} %
\newcommand{\numtopprojsast}{958\xspace}
\newcommand{\numtopprojecsastperc}{19\%\xspace} %

\newcommand{\numconfirmedsecdefects}{302\xspace}
\newcommand{\numcve}{2\xspace}
\newcommand{\percentallconfirmeddefects}{53\%\xspace} %
\newcommand{\numreposwithdefect}{97\xspace}
\newcommand{\numreposwithsecdefect}{85\xspace}
\newcommand{\percentreposwithdefect}{64\%\xspace}
\newcommand{\percentreposwithsecdefect}{56\%\xspace}
\newcommand{\numdefectsdisclosed}{586\xspace}
\newcommand{\numsecdefectsdisclosed}{433\xspace}
\newcommand{\percsecdefectsconfirmed}{70\%\xspace} %
\newcommand{\numprraised}{163\xspace}
\newcommand{\numprmerged}{104\xspace}
\newcommand{\numerrors}{772\xspace}
\newcommand{\numwarnings}{2,286\xspace}
\newcommand{\numreposfortpfp}{123\xspace} %
\newcommand{\numjuliettpbycodeql}{7,904\xspace}

\newcommand{\numemailsent}{104\xspace}

\newcommand{\numrandomsampling}{20\xspace}
\newcommand{\numrandomsamplingrepos}{30\xspace}
\newcommand{\randomsamplingmaxnumwarnings}{20\xspace}

\newcommand{\codeqlfpperc}{34\%\xspace}
\newcommand{\codeqltpperc}{66\%\xspace}
\newcommand{\codeqlnumwarnsampledfortpfp}{1577\xspace}
\newcommand{\codeqlnumtp}{1039\xspace}
\newcommand{\codeqlnumfp}{538\xspace}

\newcommand{\numcodeqlworkflowpullquests}{129\xspace}
\newcommand{\nummergedworkflows}{37\xspace}
\newcommand{\nummergedworkflowsperc}{29\%\xspace}
\newcommand{\percmergedwfactiverepos}{71\%\xspace}

\newcommand{\numhardware}{18}
\newcommand{\numdevicedriver}{10}
\newcommand{\numnetworks}{54}
\newcommand{\numdbaccess}{8}
\newcommand{\numfilesys}{5}
\newcommand{\numparsing}{10}
\newcommand{\numlanguagesupport}{33}
\newcommand{\numUI}{14}
\newcommand{\numembedapp}{32}
\newcommand{\numOS}{42} %
\newcommand{\nummemmanage}{4}
\newcommand{\numgenlib}{22}

\newcommand{\numother}{6}

%% file: tables/mainresults.tex
\begin{table*}[t]
\centering
\footnotesize
\caption{
  Summary of~\codeql{} results: setup, raw data, manual analysis, responsible disclosure, and workflow integration.
}
\label{tab:codeql-results}

\setlength{\tabcolsep}{5pt} %

\begin{minipage}[t]{0.48\textwidth}
\vspace{0pt} %
\centering
\begin{tabular*}{\linewidth}{@{\extracolsep{\fill}}lr@{}}
\toprule
\textbf{Number of ...} & \textbf{Value} \\
\midrule
\multicolumn{2}{c}{\textbf{Setup}}  \\
\midrule
Repos in dataset                             & \numrepos            \\
Repos built                                  & \numreposbuilt       \\
Repos analyzed                               & \numreposanalyzed    \\
\midrule
\multicolumn{2}{c}{\textbf{\codeql{} Results}}  \\
\midrule
Errors reported                              & \numerrors           \\
Warnings reported                            & \numwarnings         \\
\midrule
\multicolumn{2}{c}{\textbf{Manual Analysis}}   \\
\midrule
Defects discovered                           & \numalldefects       \\
Repos with defects                           & \numreposwithdefect (\percentreposwithdefect{})  \\
Security defects                             & \numsecuritydefects  \\
Repos with security defects                  & \numreposwithsecdefect (\percentreposwithsecdefect{})  \\
\bottomrule
\end{tabular*}
\end{minipage}%
\quad
\begin{minipage}[t]{0.48\textwidth}
\vspace{0pt} %
\centering
\begin{tabular*}{\linewidth}{@{\extracolsep{\fill}}lr@{}}
\toprule
\textbf{Number of ...} & \textbf{Value} \\
\midrule
\multicolumn{2}{c}{\textbf{Responsible Disclosure}}   \\
\midrule
Defects disclosed                            & \numdefectsdisclosed     \\
Defects confirmed                            & \numallconfirmeddefects  \\
Security defects disclosed                   & \numsecdefectsdisclosed  \\
Security defects confirmed                   & \numconfirmedsecdefects  \\
Patch PRs submitted                          & \numprraised         \\
Patch PRs merged (\ie{}~accepted)            & \numprmerged         \\
CVEs issued                                  & \numcve              \\
\midrule
\multicolumn{2}{c}{\textbf{\codeql{}~\ac{SAST} Workflow}}   \\
\midrule
Workflow PRs submitted                       & \numcodeqlworkflowpullquests         \\
Workflow PRs merged                          & \nummergedworkflows (\percmergedwfactiverepos{} Active) \\
                                             & and \nummergedworkflowsperc{} Total) \\
\bottomrule
\end{tabular*}
\end{minipage}
\end{table*}

%% file: fig/cdfbugsperrepo.tex
\begin{wrapfigure}{R}{0.45\textwidth}
    \centering
    \includegraphics[scale=0.55]{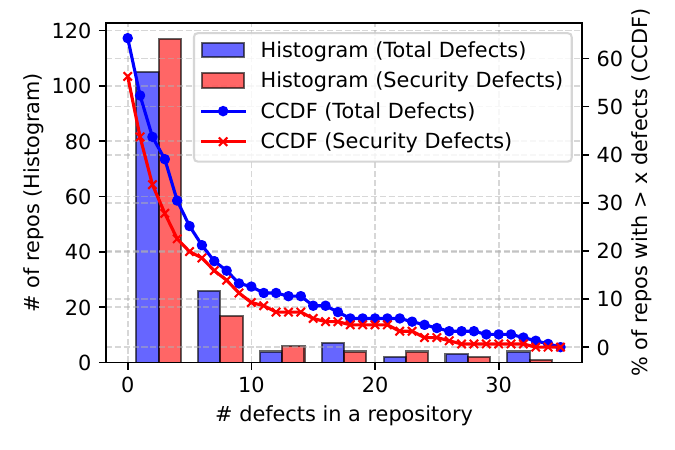}
    \caption{
    Histograms (left axis) and CCDFs (right axis) of \# of total and security-relevant defects in a repository.
    The CCDF values at 0 defects indicate that defects are discovered in \percentreposwithdefect of repositories, and security-relevant defects are discovered in \percentreposwithsecdefect of repositories.
    }
    \label{fig:cdf-bugs-per-repo}
\end{wrapfigure}

%% file: tables/topreposwithdefects.tex
\begin{wraptable}{r}{0.63\textwidth}
\centering
\footnotesize
\caption{
  Top-5~\ac{EMBOSS} repositories by number of total defects found.
  }
\label{tab:topreposwithdefects}
\begin{tabular}{@{}cccc@{}}
\toprule
\textbf{Repo}         & \textbf{Criti. Score} & \textbf{\# Total} & \textbf{\# Security} \\ \midrule
apache/nuttx          & 0.69                       & 35 & 24                       \\
contiki-ng/contiki-ng & 0.67                       & 34  & 24                      \\
raysan5/raylib        & 0.70                       & 33  & 33                      \\
ARMmbed/mbed-os       & 0.72                       & 32  & 22                      \\
openlgtv/epk2extract  & 0.45                       & 29  & 27                      \\ \bottomrule
\end{tabular}
\end{wraptable}

%% file: fig/codeqlqueryfig.tex
\begin{wrapfigure}{r}{0.63\textwidth}
    \centering
    \begin{tikzpicture}
    \begin{axis}[
    xbar,
    width=5.5cm,
    xmax=170,
    bar width=0.2cm,
    height=4.5cm,
    enlargelimits=0.1,
    legend style={at={(0.5,-0.15)},
    anchor=north,legend columns=-1},
    xlabel={\# security-relevant defects},
    symbolic y coords={
        cpp/toctou-race-condition,
        cpp/incorrect-allocation-error-handling,        
        cpp/overflowing-snprintf,
        cpp/offset-use-before-range-check,
        cpp/uninitialized-local,    
        cpp/wrong-type-format-argument,
        cpp/unbounded-write,        
        cpp/uncontrolled-allocation-size,
        cpp/missing-check-scanf,
        cpp/inconsistent-null-check,        
    },
    ytick=data,
    yticklabel style={font=\scriptsize},
    nodes near coords={
        \pgfplotspointmeta
    },
    nodes near coords style={font=\scriptsize, anchor=west},
    point meta=explicit symbolic,
    nodes near coords align={horizontal},
    ]
    \addplot table[x=bugs, y=rule, meta=perc] {data/bugs-per-type.dat};
    \end{axis}
    \end{tikzpicture}    
    \caption{
    Top-10 CodeQL queries by security-relevant defects found.}
    \label{fig:bugs-per-rule}
\end{wrapfigure}

%% file: fig/cdfbugseverity.tex
\begin{wrapfigure}[13]{r}{0.4\textwidth}
    \centering
    \vspace{-0.3cm}
    \begin{tikzpicture}
    \begin{axis}[
      xlabel={criticality score},
      ylabel={\% of security defects},
      xmin=0, xmax=1,
      ymin=0, ymax=1,
      grid=major,
      grid style={dashed},
      width=5cm,
      yticklabel={\pgfmathparse{\tick*100}\pgfmathprintnumber{\pgfmathresult}\%},
    ]
    
    \addplot table[x=score, y=ccdf]{data/bug-severity-cdf.dat};
    \end{axis}
    \end{tikzpicture}
    \caption{
    CCDF of the severity of security defects.
    }
    \label{fig:cdf-bug-severity}
\end{wrapfigure}

%% file: listings/unboundedwrite.tex
\begin{minipage}{0.46\textwidth}
\captionsetup{type=listing}
\begin{minted}[highlightlines={}]{c}
// mbedtls/programs/ssl/ssl_mail_client.c
int main(int argc, char *argv[]) {
  unsigned char buf[1024];
  char *q = strchr(argv[1], '=') + 1;
  opt.mail_from = q; @\textcolor{purple}{\faWarning}@
  len = @\textcolor{red}{\faBomb}@sprintf((char *) buf, "MAIL FROM:<%
  ...
}
\end{minted} 
\caption{Unbounded \code{sprintf} (\textcolor{red}{\faBomb}) formats attacker‑controlled \code{opt.mail_from} (\textcolor{purple}{\faWarning}) into the fixed‑size stack buffer \code{buf}, enabling a classic stack buffer overflow.
(Code is simplified for clarity.)}
\label{lst:unbounded-write-mbedtls}
\end{minipage}

%% file: listings/useafterfreelisting.tex
\begin{minipage}{0.46\textwidth}
\captionsetup{type=listing}
    \begin{minted}[highlightlines={}]{c}
    // apache/nuttx/drivers/sensors/apds9960.c
    ret = register_driver(devpath, &g_apds9960_fops, 0666, priv);
    if (ret < 0) {
      snerr("ERROR: Failed to register driver: %
      kmm_free(priv)@\textcolor{purple}{\faWarning}@;
    }
    @\textcolor{red}{\faBomb}@priv->config->irq_attach(priv->config, apds9960_int_handler, priv);
    \end{minted}

    \caption{
    The memory pointed by~\code{priv} is freed (\textcolor{purple}{\faWarning}) inside the~\code{if} condition.
    It is accessed later on, resulting in use-after-free (\textcolor{red}{\faBomb}).
    }
    \label{lst:useafterfreeinsensor}
\end{minipage}

%% file: fig/nonsecbugsperrule.tex
\begin{wrapfigure}{r}{0.6\textwidth}
    \centering
    \begin{tikzpicture}
    \begin{axis}[
    xbar,
    width=5.5cm,
    xmax=45,
    bar width=0.2cm,
    height=4.5cm,
    enlargelimits=0.1,
    legend style={at={(0.5,-0.15)},
    anchor=north,legend columns=-1},
    xlabel={\# non-security-relevant defects},
    symbolic y coords={
cpp/unsigned-comparison-zero,
cpp/nested-loops-with-same-variable,
cpp/comparison-precedence,
cpp/non-member-const-no-effect,
cpp/virtual-call-in-constructor,
cpp/duplicate-include-guard,
cpp/implicit-function-declaration,
cpp/constant-comparison,
cpp/stack-address-escape,
cpp/missing-return,
},
    ytick=data,
    yticklabel style={font=\scriptsize},
    nodes near coords={
        \pgfplotspointmeta
    },
    nodes near coords style={font=\scriptsize, anchor=west},
    point meta=explicit symbolic,
    nodes near coords align={horizontal},
    ]
    \addplot table[x=bugs, y=rule, meta=perc] {data/non-sec-bugs-per-type.dat};
    \end{axis}
    \end{tikzpicture}    
    \caption{
    Top-10 CodeQL queries by non-security defects found. %
    }
    \label{fig:non-sec-bugs-per-rule}
\end{wrapfigure}

%% file: fig/bugssplit.tex
\pgfplotstableread{data/bugspercategory.dat}\mydata

\begin{figure*}[t]
\centering
\footnotesize
\begin{tikzpicture}%
\begin{axis}[
    axis background/.style={fill=gray!10},
    ybar stacked,
    x tick label style={
		/pgf/number format/1000 sep=},
	bar width=16pt,
	width=14cm,
	height=5cm,
	nodes near coords,
	ymin=1,
    enlarge x limits=0.03,
    enlarge y limits=0.03,
    legend style={at={(0.015,0.95)},
      anchor=north west,legend columns=-1, font=\small},
    ylabel={Number of defects},
    symbolic x coords={HAL(5), DD(4), NET(20), DAL(3), FS(1), PAR(3), LS(15), UI(7), APP(12), OS(16), MML(1), GPL(6), OT(4)},
    xtick=data,
    every node near coord/.style={text=black, font=\bfseries},
    cycle list={fill=red!55,fill=red!15,fill=red!5,fill=purple!50}
    ]
\pgfplotsinvokeforeach{ss,nss}{ %
  \addplot table[x=embosstype,y=#1,text=black] {\mydata};
}
\legend{\strut Security, \strut Non-security}
\end{axis}
\end{tikzpicture}
    \caption{
    Number of defects found in repositories of each category (defined in \tbl{tab:EMBOSSCorpus-Summary}).
    On x-axis, numbers show \# repositories with $\geq1$ defect.
    }
    \label{fig:vultypesinemboss}%
    \vspace{-0.1cm}
\end{figure*}

%% file: fig/falsepositivesperrulesfig.tex
\begin{figure*}[h]
\footnotesize
\centering
\begin{subfigure}[t]{0.33\textwidth}
    \centering
    \begin{tikzpicture}
    \begin{axis}[
      xlabel={False positive rate},
      ylabel={\% of rules},
      xmin=0, xmax=1,
      ymin=0, ymax=1,
      grid=major,
      grid style={dashed},
      xticklabel={\pgfmathparse{\tick*100}\pgfmathprintnumber{\pgfmathresult}\%},
      yticklabel={\pgfmathparse{\tick*100}\pgfmathprintnumber{\pgfmathresult}\%}, %
      width=5cm,
      height=4cm,
      legend pos=south east,
    ]
    \addplot table [x=percent,y=fp] {data/rule-tp-fp.dat};

    \end{axis}
    \end{tikzpicture}
    \caption{\parbox[t]{3.5cm}{CDF of the false positive rates of rules}}
    \label{fig:cdf-tp-fp}
\end{subfigure}
~
\begin{subfigure}[t]{0.33\textwidth}
    \centering
    \begin{tikzpicture}
    \begin{axis}[
      xlabel={False positive rate},
      ylabel={\% of repos},
      xmin=0, xmax=1,
      ymin=0, ymax=1,
      grid=major,
      grid style={dashed},
      xticklabel={\pgfmathparse{\tick*100}\pgfmathprintnumber{\pgfmathresult}\%},
      yticklabel={\pgfmathparse{\tick*100}\pgfmathprintnumber{\pgfmathresult}\%}, %
      width=5cm,
      height=4cm,
    ]
    \addplot table [x=fp,y=cdf] {data/fp-perc-per-repo-cdf.dat};

    \end{axis}
    \end{tikzpicture}
    \caption{\parbox[t]{3.5cm}{CDF of the false positive rate vs. \% of repos} }
    \label{fig:fp-perc-per-repo-cdf}
\end{subfigure}
~
\begin{subfigure}[t]{0.33\textwidth}
    \centering
    \begin{tikzpicture}
    \begin{axis}[
      xlabel={\# of false positives},
      ylabel={\% of repos},
      xmin=0, xmax=100,
      ymin=0, ymax=1,
      grid=major,
      grid style={dashed},
      xmode=log,
      yticklabel={\pgfmathparse{\tick*100}\pgfmathprintnumber{\pgfmathresult}\%}, %
      width=5cm,
      height=4cm,
    ]
    \addplot table [x=fp,y=cdf] {data/fp-num-per-repo-cdf.dat};

    \end{axis}
    \end{tikzpicture}
    \caption{\parbox[t]{3.5cm}{CDF of \# of false positives vs. \% of repos}}
    \label{fig:fp-num-per-repo-cdf}
\end{subfigure}
\caption{
CDF of the false positive rates of rules, along with
CDFs of the rate and number of false positives vs. percentage of repositories (repos).
}
\end{figure*}

%% file: newappendix-issta2025.tex
\appendix

\section{Outline of appendices} \label{apdx:outline}

\begin{itemize}
    \item \cref{apdx:outline}: This outline.
    \item \cref{apdx:survey-questions}: Developer survey instrument (cf. \cref{sec:Theme1-RQ1-Methods-DeveloperSurvey}).
    \item \cref{apdx:selection-of-sast-tools}: Selection of SAST tools (cf. \cref{subsec:selectionofsast}).
    \item \cref{apdx:ExamplePR}: Example PR opened to propose a CodeQL workflow (cf. \cref{sec:rq-codeql-wf}).
    \item \cref{apdx:workflowbackground}: Further details and example GitHub CI pipeline (cf. \cref{sec:Background-CI}).
    \item \cref{apdx:MoreAnalysisResults}: Further results of applying CodeQL to the EMBOSS dataset (cf. \cref{sec:codeql-effectiveness}).
    \item \cref{apdx:gcceffectiveness}: Evaluation of compiler flags in lieu of CodeQL (cf. \cref{subsubsec:developersurvey}).
\end{itemize}

\section{Developer survey questions}
\label{apdx:survey-questions}
\begin{enumerate}
    \item What is the name of the open-source project that brought you to this survey (owner/repo, e.g. torvalds/linux)?
    \item What roles do you play in the project? (select all that apply)
        \begin{itemize}
            \item Developer (you implement new features)
            \item Tester (you validate that new features work properly and that regressions are avoided)
            \item Maintainer (you give feedback on issues, fix defects, and review and merge PRs)
            \item Owner (e.g., you help determine the future path of the project / steering member)
        \end{itemize}
    \item How long have you been involved in the project?
        \begin{itemize}
            \item Less than 1 year
            \item 1-3 years
            \item 4-6 years
            \item More than 6 years
        \end{itemize}
    \item What level of risk do you perceive if a functionality defect is present in this project?
        \begin{itemize}
            \item Critical
            \item High
            \item Medium
            \item Low
        \end{itemize}
    \item What level of risk do you perceive if a security vulnerability is present in this project?
        \begin{itemize}
            \item Critical
            \item High
            \item Medium
            \item Low
        \end{itemize}
    \item What is your experience with static analysis tools (SAST) such as CodeQL, Flawfinder, CodeChecker, cppcheck, clang-tidy, Clang Static Analyzer, and `gcc -Wall -Werror`?
        \begin{itemize}
            \item I use them regularly on this project
            \item I use them regularly on other projects but not this project
            \item I have read about them but not used them
            \item I tried them out but stopped using them
            \item I have never used them
        \end{itemize}
    \item Do any corporations or other entities support this software? (check all that apply)
        \begin{itemize}
            \item Yes, financially (e.g. donations)
            \item Yes, technically (e.g. engineers)
            \item Yes, through infrastructure (e.g., servers)
            \item No
        \end{itemize}
    \item Do you know that SAST tools can be integrated into GitHub Workflows and can be configured to run on various events (e.g. when a commit is pushed, when a pull request is opened, etc)?
        \begin{itemize}
            \item Yes
            \item No
        \end{itemize}
    \item What SAST tools are you using in GitHub Workflows for this project? (select all that apply)
        \begin{itemize}
            \item CodeQL
            \item FlawFinder
            \item CodeChecker
            \item cppcheck
            \item clang-tidy
            \item Clang Static Analyzer
            \item gcc with -Wall / -Wextra / -Werror
            \item Other
            \item No SAST tools
        \end{itemize}
    \item Are you running any SAST tool outside the GitHub workflow, e.g. in Makefile, CMakeLists.txt, git-hooks, etc.?
        \begin{itemize}
            \item Yes
            \item No
        \end{itemize}
    \item What SAST tools are you using outside of GitHub Workflows? (select all that apply)
        \begin{itemize}
            \item CodeQL
            \item FlawFinder
            \item CodeChecker
            \item cppcheck
            \item clang-tidy
            \item Clang Static Analyzer
            \item gcc with -Wall / -Wextra / -Werror
            \item Other
            \item No SAST tools
        \end{itemize}
    \item Why didn't you incorporate these SAST tools into a GitHub Workflow?
        \begin{itemize}
            \item We use them in another CI (e.g. Travis CI).
            \item There is no GitHub Action for the SAST tool we use.
            \item I don't have time / Too much work
            \item Other
        \end{itemize}
    \item Why don't you use SAST tools for this project? (select all that apply)
        \begin{itemize}
            \item I didn't bother, it's not a mission-critical project
            \item Difficulty in configuration
            \item Too many false positives
            \item Other
        \end{itemize}
    \item This project is embedded software. Does that affect your use of SAST?
        \begin{itemize}
            \item No
            \item Yes, please explain.
        \end{itemize}
    \item Is there anything else you would like to tell us about your experiences with SAST related to this project or other embedded projects?
\end{enumerate}

\section{Selection of SAST tools}
\label{apdx:selection-of-sast-tools}
Our goal is to find~\ac{SAST} tools that can be readily used on the collected GitHub repositories.
Given that GitHub Actions are expected to be stable, easy to use, and can be seamlessly integrated into repositories,
we used GitHub Marketplace and found GitHub Actions designed for~\ac{SAST} purposes.~\footnote{The query is category=security\&type=actions\&query=``C C++'' and category=code-quality\&type=actions\&query=``C C++''.}
We manually filtered out pre-release Actions due to their instability and/or lack of documentation. %
There were 6 commercial~\ac{SAST} tools, which we omitted as they require purchases of licenses or subscriptions and place restrictions on scientific publications.

This resulted in a total of \numsast GitHub Actions using various~\ac{SAST} tools as shown in~\tbl{tbl:sast-tools}.
Nine of these Actions are plug-and-play, meaning they do not need any repository-specific configuration. %

Subsequently, for each plug-and-play~\ac{SAST} tool, we picked the most popular Action implementing it.
For instance, for~\code{flawfinder}, we picked~\code{david-a-wheeler/flawfinder} --- this repository has the most stars among the Actions offering this tool.
This resulted in the~\ac{SAST} tools selected in~\tbl{tab:SASTToolComparison}.

{
\begin{table*}[ht]
\caption{
  List of ``usable'' SAST GitHub Actions.
  These are GitHub Actions that perform \ac{SAST} on C/C++ repositories, not including pre-release or commercial tools.
  }
\label{tbl:sast-tools}
\centering
\scriptsize
\begin{tabular}{@{}lclc@{}}
\toprule
\textbf{Name of GitHub Action} &
  \textbf{Plug-and-play?} &
  \textbf{Underlying tool(s)} &
  \textbf{Juliet Test Suite results} \\ \midrule
    \multicolumn{4}{c}{\textbf{\textit{From Well-established Organizations}}} \\
\rowcolor[HTML]{EFEFEF} 
github/codeql-action~\cite{github_inc_codeql_2023} &
  Yes &
  \codeql{} &
  \numjuliettpbycodeql true positives \\
cpp-linter/cpp-linter-action~\cite{cpp-linter_cc_2023} &
  Yes &
  clang-format, clang-tidy &
  Not finished in 6 hours \\
\rowcolor[HTML]{EFEFEF} 
trunk-io/trunk-action~\cite{trunk-io_trunkio_2023} &
  \begin{tabular}[c]{@{}c@{}}No\\ (Bazel/CMake projects required)\end{tabular} &
  \begin{tabular}[c]{@{}l@{}}clang-format, clang-tidy,\\ include-what-you-use,\\ pragma-once\end{tabular} &
  N/A \\
Frama-C/github-action-eva-sarif~\cite{frama-c_frama-ceva_2023} &
  \begin{tabular}[c]{@{}c@{}}No\\ (Frama-C Makefile required)\end{tabular} &
  Frama-C &
  N/A \\ \midrule
  \multicolumn{4}{c}{\textit{\textbf{From Independent Developers}}} \\
\rowcolor[HTML]{EFEFEF} 
IvanKuchin/SAST~\cite{ivankuchin_cc_2021} &
  \cellcolor[HTML]{EFEFEF} &
  \cellcolor[HTML]{EFEFEF} &
  \cellcolor[HTML]{EFEFEF} \\
\rowcolor[HTML]{EFEFEF} 
deep5050/flawfinder-action~\cite{pal_flawfinder-action_2022} &
  \cellcolor[HTML]{EFEFEF} &
  \cellcolor[HTML]{EFEFEF} &
  \cellcolor[HTML]{EFEFEF} \\
\rowcolor[HTML]{EFEFEF} 
david-a-wheeler/flawfinder~\cite{wheeler2006flawfinder} &
  \multirow{-3}{*}{\cellcolor[HTML]{EFEFEF}Yes} &
  \multirow{-3}{*}{\cellcolor[HTML]{EFEFEF}flawfinder} &
  \multirow{-3}{*}{\cellcolor[HTML]{EFEFEF}Error} \\
Syndelis/cpp-linter-cached-action~\cite{lemos_cc_2023} &
  Yes &
  clang-format, clang-tidy &
  N/A \\
\rowcolor[HTML]{EFEFEF} 
deep5050/cppcheck-action~\cite{pal_deep5050cppcheck-action_2023} &
  \cellcolor[HTML]{EFEFEF} &
  \cellcolor[HTML]{EFEFEF} &
  \cellcolor[HTML]{EFEFEF} \\
\rowcolor[HTML]{EFEFEF} 
Konstantin343/cppcheck-annotation-action~\cite{konstantin343_cppcheck_2022} &
  \multirow{-2}{*}{\cellcolor[HTML]{EFEFEF}Yes} &
  \multirow{-2}{*}{\cellcolor[HTML]{EFEFEF}cppcheck} &
  \multirow{-2}{*}{\cellcolor[HTML]{EFEFEF}Not finished in 6 hours} \\
JacobDomagala/StaticAnalysis~\cite{domagala_static_2023} &
  Yes &
  cppcheck, clang-tidy &
  Error \\
\rowcolor[HTML]{EFEFEF} 
whisperity/codechecker-analysis-action~\cite{whisperity_codechecker_2023} &
  \begin{tabular}[c]{@{}c@{}}No\\ (Compilation DB required)\end{tabular} &
  clang &
  N/A \\ \bottomrule
\end{tabular}
\end{table*}
}

\input{pullrequestinfo}

\section{\acf{CI} Pipeline}
\label{apdx:workflowbackground}
A~\ac{CI} pipeline is event-driven: upon a triggering event, the CI framework executes a sequence of steps.
For example,~\lst{lst:workflow} shows an example of a GitHub Workflow triggered on a push (\encircle{\textbf{1}}) to the underlying repository.
The Workflow has five steps (\encircle{\textbf{2}} - \encircle{\textbf{6}}).
The first two steps,~\ie{} Build Project (\encircle{\textbf{2}}) and Test Project (\encircle{\textbf{3}}), will build and run the tests on the project with newly pushed changes.
The last three steps (\encircle{\textbf{4}} - \encircle{\textbf{6}}) are related to running~\codeql{} (a~\ac{SAST} tool) on the repository.

\begin{listing}[h]
\begin{minted}[breaklines, mathescape, escapeinside=||, fontsize=\footnotesize{}]{YAML}
name: MyWorkflow

on:
  # Workflow triggers on push 
  push |\encircle{\textbf{1}}|
    
steps: |$\leftarrow$ \textcolor{blue}{\textbf{Steps}}|
    # The following steps are executed sequentially 

    ### Build and functional test

    - name: Build Project  |\encircle{\textbf{2}}|
      uses: actions/cmake-action
    
    - name: Test Project  |\encircle{\textbf{3}}|
      run: ./test.sh
      
    ### Execute CodeQL SAST
    
    # Initialize
    - name: Initialize CodeQL  |\encircle{\textbf{4}}|
      uses: github/codeql-action/init@v2
      
    # Build the Code
    - name: Autobuild  |\encircle{\textbf{5}}|
      uses: github/codeql-action/autobuild@v2
      
    # Run the analysis
    - name: Perform CodeQL Analysis  |\encircle{\textbf{6}}|
      uses: github/codeql-action/analyze@v2
\end{minted}
\caption{
    Snippet of a GitHub Workflow (\ie{} a~\code{YML} file) that builds, tests, and runs~\codeql{} on the underlying repository.
    The various Actions are taken from the GitHub CI marketplace of prebuilt actions~\cite{githubmarketplace}.
    }
\label{lst:workflow}
\end{listing}

\section{Further SAST Analysis Results with CodeQL} \label{apdx:MoreAnalysisResults}

\subsection{Common Types of Security Defects}
\label{apdx:commontypesofsecuritydefects}
We discuss three major types of defects and corresponding rules:
\begin{itemize}[leftmargin=*]
    \item \code{cpp/inconsistent-null-check}: This rule identifies cases where the return value of a function is not checked for~\code{NULL}, despite most other calls to the same function performing such a check. Developers should consistently validate return values that may be~\code{NULL} to prevent potential null pointer dereferences. This rule flagged 135 such instances.
This rule detected 135 such instances.
\lst{lst:inconsistent-null-check} shows an instance of this issue from the~\code{ARMmbed/mbed-os} repository.

    \item \code{cpp/uncontrolled-allocation-size}: This rule detects cases where the size argument of a memory allocation function (\eg~\code{malloc}) is computed through integer arithmetic involving potentially untrusted input (\eg~user input). If the input takes on large values, an integer overflow~\cite{dietz2015understanding} may occur, resulting in an allocation size significantly smaller than intended. Subsequent buffer accesses may lead to out-of-bounds reads or writes. This rule identified 49 such instances.
\lst{lst:uncontrolled-allocation-size} shows an instance of this defect in the~\code{embox/embox} repository.

    \item \code{cpp/unbounded-write}: This rule detects out-of-bound write vulnerabilities.
Specifically, this includes analysis of potentially dangerous function calls (\eg{}~\code{strcpy}, \code{sscanf}) to check whether these are used properly with valid arguments.
This rule detected 47 vulnerabilities of potential buffer overflow.
\lst{lst:unbounded-write} shows an instance of this vulnerability in the \\
~\code{aws}/\code{aws-iot-device-sdk-embedded-C} repository.

\end{itemize}

\begin{listing}[htb]
\begin{minted}[highlightlines={}]{c}
// mbed-os/connectivity/FEATURE_BLE/libraries/cordio_stack/ble-host/sources/stack/att/att_eatt.c
static uint8_t eattL2cCocAcceptCback(dmConnId_t connId, uint8_t numChans)
{
  eattConnCb_t *pCcb = eattGetConnCb(connId)@\textcolor{purple}{\faWarning}@;

  if ((pCcb->state@\textcolor{red}{\faBomb}@ == EATT_CONN_STATE_INITIATING) ...
\end{minted}

\caption{
The return value of \code{eattGetConnCb} (\textcolor{purple}{\faWarning}) is not checked for null, possibly leading to null pointer dereference (\textcolor{red}{\faBomb}).
}
\label{lst:inconsistent-null-check}
\end{listing}

    \begin{listing}[htb]
    \begin{minted}[highlightlines={}]{c}
// embox/src/cmds/testing/block_dev/block_dev_test.c
static int block_dev_test(struct block_dev *bdev, uint64_t s_block, uint64_t n_blocks, uint64_t m_blocks@\textcolor{purple}{\faWarning}@) {
    size_t blk_sz;
    int8_t *read_buf, *write_buf;

    blk_sz = bdev->block_size;
    if (blk_sz == 0) {
        return -1;
    }
    ...
    read_buf = malloc(blk_sz * m_blocks@\textcolor{red}{\faBomb}@);
    write_buf = malloc(blk_sz * m_blocks@\textcolor{red}{\faBomb}@);
    ...
}
    \end{minted} 
    \caption{The variable \code{m_blocks} is tainted (\textcolor{purple}{\faWarning}), \ie derived from user inputs, whose value may be huge. The multiplication that calculates the \code{malloc} size may overflow (\textcolor{red}{\faBomb}), leading to the allocation size being considerably smaller than expected.}
    \label{lst:uncontrolled-allocation-size}
    \end{listing}

    \begin{listing}[htb]
\begin{minted}[highlightlines={}]{c}
// metrics_collector.c
MetricsCollectorStatus_t GetNetworkInferfaceInfo(
    char ( *pOutNetworkInterfaceNames )[16]@\textcolor{blue}{\faInfo}@,
    uint32_t * pOutNetworkInterfaceAddresses,
    size_t bufferLength,
    size_t * pOutNumNetworkInterfaces )
{
  char lineBuffer[ MAX_LINE_LENGTH ];
  ...
  while((*pOutNumNetworkInterfaces < bufferLength)
    && (fgets(&(lineBuffer[0])@\textcolor{purple}{\faWarning}@,
        MAX_LINE_LENGTH, fileHandle) != NULL))
  {
    filledVariables = sscanf(
      lineBuffer@\textcolor{purple}{\faWarning}@,
      "%
      &ipPart1,
      &ipPart2,
      &ipPart3,
      &ipPart4,
      pOutNetworkInterfaceNames[
        *pOutNumNetworkInterfaces]@\textcolor{red}{\faBomb}@);
    ...
  }
}
    \end{minted} 
    \caption{The content of \code{lineBuffer} is read from a file via \code{fgets}.
    A part of it is then copied to the \code{char} array \code{pOutNetworkInterfaceNames[*pOutNumNetworkInterfaces]}, whose size is 16 bytes (\textcolor{blue}{\faInfo}), via \code{sscanf} with the format specifier ``\%s''. Since \code{lineBuffer} is tainted (\textcolor{purple}{\faWarning}), the content copied to the \code{char} array may be strictly longer than 15 bytes (one byte is needed for the null terminator). Thus, the \code{sscanf} call may lead to buffer overwrite (\textcolor{red}{\faBomb}).}
    \label{lst:unbounded-write}
    \end{listing}

\subsection{Defect Density Per Category}
\label{apdx:defect_density}
\cref{fig:defect_density} shows the defect density per number of repositories and KLOC across EMBOSS types.

\def\mathdefault#1{#1}
\begin{figure*}
    \centering
    \scalebox{0.3}{\input{fig/normalized_defects.pgf}}
    \caption{Defect density per number of repositories and KLOC across EMBOSS types}
    \label{fig:defect_density}
\end{figure*}

\subsection{Rules contributing to False Positives}
\label{apdx:commonfprules}

The following are the top four queries contributing to false positives:

\begin{itemize}[leftmargin=*]
    \item cpp/uninitialized-local. Dataflow analysis of CodeQL is not path-sensitive. Some variables may not be initialized in all paths. However, when a variable is used, certain path conditions hold, under which it can be proved that the variable must have been initialized. \lst{lst:uninit-fp} shows an example.
    \item cpp/missing-check-scanf. Developers can use \code{switch case} statements (instead of \code{if} statements) to check the return value of \code{scanf} calls. These are valid checks but not detected by CodeQL.
    \item cpp/suspicious-pointer-scaling and cpp/suspicious-pointer-scaling-void. These rules detect risky pointer arithmetic operations. However, pointer casts, and type-punning are pretty common and unavoidable in low-level embedded system code.
    \item cpp/unbounded-write. \code{strcpy} is safe if the destination must be large enough. For example, developers can first use \code{strlen} to calculate the length of the source string, allocate enough memory for the destination string, and then call \code{strcpy}.
\end{itemize}

\tbl{tab:DetailedCodeQLQueryPerformance} provides detailed results across different~\codeql{} rules.

    \begin{listing}[htb]
    \begin{minted}[highlightlines={},linenos]{c}
// zephyr/drivers/timer/nrf_rtc_timer.c
static struct z_nrf_rtc_timer_chan_data cc_data[CHAN_COUNT];
static void process_channel(int32_t chan)
{
  void *user_context;
  uint64_t curr_time;
  uint64_t expire_time;
  z_nrf_rtc_timer_compare_handler_t handler = NULL;

  curr_time = z_nrf_rtc_timer_read();
  ...
  expire_time = cc_data[chan].target_time;
  if (curr_time >= expire_time) {
    handler = cc_data[chan].callback;
    user_context = cc_data[chan].user_context;
    ...
  }
  ...  
  if (handler) {
    handler(chan, expire_time, user_context@\textcolor{purple}{\faWarning}@);
  }
}
    \end{minted} 
    \caption{CodeQL reports the variable \code{user_context} is used uninitialized (\textcolor{purple}{\faWarning}) because it is not initialized in all paths. However, when it is used, \code{handle} must be non-null, which means that the assignment in line 14 must have been executed (since this is the only place where \code{handle} can get a non-null value). It follows that line 15 must also have been executed. Consequently, the variable \code{user_context} must have been initialized in line 20.}
    \label{lst:uninit-fp}
    \end{listing}

\input{tables/rulewisefps}

\section{Effectiveness of Stringent Compiler Flags Instead of CodeQL}
\label{apdx:gcceffectiveness}
\ac{EMBOSS} developers often use strict compiler flags/warnings instead of~\ac{SAST} tools.
We evaluated the effectiveness of these flags in finding the defects detected by~\codeql{}.
We used security bug test case files from the~\codeql{} repository for this experiment.
These are simple test cases (< 10 lines), each containing an obvious security issue,~\eg{}
  passing an invalid pointer types to a function call.
We selected test cases to cover all~\numdefecttypes{} of the identified defect types and compiled them using the latest version of~\code{gcc}, \ie{}~\latestgccversion{}, with strict warnings (~\code{-Wall, -Wextra -Werror}).
This configuration of \code{gcc} found issues in only~\numdefecttypesfound{} (\numdefecttypesfoundperc{}) defect types as shown in~\tbl{tab:gcc-detection-of-codeql-queries}.
~\code{gcc} was able to find certain simple security issues, such as direct use of~\code{strcpy}.
However, it did not find more complex ones related to code quality, such as inconsistent null check.
\textit{Our results indicate that the current~\ac{EMBOSS} practice of reliance on~\code{gcc} warnings is inadequate}.

\topcaption{GCC -Wall Detection of CodeQL Queries}
\label{tab:gcc-detection-of-codeql-queries}

\tablefirsthead{ %
  \toprule
  \textbf{Issue} & \textbf{Detected} \\
  \midrule
}

\tablehead{
  \multicolumn{2}{c}{\textit{Continued from previous page}} \\
  \midrule
  \textbf{Issue} & \textbf{Detected} \\
  \midrule
}

\tablelasttail{
  \bottomrule
}

\tabletail{
  \midrule
  \multicolumn{2}{c}{\textit{Continued on next page}} \\
}

\definecolor{Gray}{gray}{0.9}

\clearpage
\begin{center}
\begin{xtabular}{l|c}
\rowcolor{Gray} cpp/alloca-in-loop & No \\
cpp/ambiguously-signed-bit-field & No \\
\rowcolor{Gray} cpp/assign-where-compare-meant & No \\
cpp/badly-bounded-write & No \\
\rowcolor{Gray} cpp/bad-strncpy-size & No \\
cpp/certificate-result-conflation & No \\
\rowcolor{Gray} cpp/cgi-xss & No \\
cpp/cleartext-transmission & No \\
\rowcolor{Gray} cpp/comma-before-misleading-indentation & No \\
cpp/command-line-injection & No \\
\rowcolor{Gray} cpp/compare-where-assign-meant & No \\
cpp/comparison-with-wider-type & Yes \\
\rowcolor{Gray} cpp/dangerous-cin & No \\
cpp/dead-code-goto & No \\
\rowcolor{Gray} cpp/double-free & No \\
cpp/external-entity-expansion & No \\
\rowcolor{Gray} cpp/HRESULT-boolean-conversion & No \\
cpp/inconsistent-loop-direction & Yes \\
\rowcolor{Gray} cpp/inconsistent-null-check & No \\
cpp/incorrect-allocation-error-handling & No \\
\rowcolor{Gray} cpp/incorrect-not-operator-usage & Yes \\
cpp/incorrect-string-type-conversion & No \\
\rowcolor{Gray} cpp/insufficient-key-size & No \\
cpp/integer-multiplication-cast-to-long & No \\
\rowcolor{Gray} cpp/logical-operator-applied-to-flag & No \\
cpp/memset-may-be-deleted & No \\
\rowcolor{Gray} cpp/missing-check-scanf & No \\
cpp/new-free-mismatch & No \\
\rowcolor{Gray} cpp/non-https-url & No \\
cpp/no-space-for-terminator & No \\
\rowcolor{Gray} cpp/offset-use-before-range-check & No \\
cpp/overflowing-snprintf & No \\
\rowcolor{Gray} cpp/path-injection & No \\
cpp/pointer-overflow-check & No \\
\rowcolor{Gray} cpp/potentially-dangerous-function & No \\
cpp/potential-system-data-exposure & No \\
\rowcolor{Gray} cpp/redundant-null-check-simple & No \\
cpp/resource-not-released-in-destructor & Yes \\
\rowcolor{Gray} cpp/return-stack-allocated-memory & Yes \\
cpp/sql-injection & No \\
\rowcolor{Gray} cpp/static-buffer-overflow & No \\
cpp/suspicious-add-sizeof & No \\
\rowcolor{Gray} cpp/suspicious-allocation-size & No \\
cpp/suspicious-pointer-scaling & No \\
\rowcolor{Gray} cpp/suspicious-sizeof & No \\
cpp/tainted-format-string & Yes \\
\rowcolor{Gray} cpp/tainted-permissions-check & No \\
cpp/toctou-race-condition & No \\
\rowcolor{Gray} cpp/uncontrolled-allocation-size & No \\
cpp/uncontrolled-arithmetic & No \\
\rowcolor{Gray} cpp/uncontrolled-process-operation & No \\
cpp/uninitialized-local & Yes \\
\rowcolor{Gray} cpp/unsafe-create-process-call & No \\
cpp/unsafe-dacl-security-descriptor & No \\
\rowcolor{Gray} cpp/unsafe-strcat & No \\
cpp/unsafe-strncat & No \\
\rowcolor{Gray} cpp/unsafe-use-of-this & Yes \\
cpp/unsigned-difference-expression-compared-zero & No \\
\rowcolor{Gray} cpp/unterminated-variadic-call & No \\
cpp/upcast-array-pointer-arithmetic & No \\
\rowcolor{Gray} cpp/use-after-free & No \\
cpp/useless-expression & Yes \\
\rowcolor{Gray} cpp/user-controlled-bypass & No \\
cpp/using-expired-stack-address & No \\
\rowcolor{Gray} cpp/weak-cryptographic-algorithm & No \\
cpp/wrong-type-format-argument & Yes \\
\rowcolor{Gray} cpp/allocation-too-small & No \\
cpp/bad-addition-overflow-check & No \\
\rowcolor{Gray} cpp/certificate-not-checked & No \\
cpp/cleartext-storage-buffer & No \\
\rowcolor{Gray} cpp/cleartext-storage-file & No \\
cpp/dangerous-function-overflow & Yes \\
\rowcolor{Gray} cpp/open-call-with-mode-argument & No \\
cpp/overrunning-write & Yes \\
\rowcolor{Gray} cpp/overrunning-write-with-float & No \\
cpp/signed-overflow-check & No \\
\rowcolor{Gray} cpp/suspicious-pointer-scaling-void & Yes \\
cpp/system-data-exposure & No \\
\rowcolor{Gray} cpp/tainted-format-string-through-global & Yes \\
cpp/too-few-arguments & Yes \\
\rowcolor{Gray} cpp/unbounded-write & Yes \\
cpp/very-likely-overrunning-write & Yes \\
\end{xtabular}
\end{center}

%% file: pullrequestinfo.tex
\section{Example pull request contributing a CodeQL workflow} \label{apdx:ExamplePR}

This appendix describes a pull request.
Each heading corresponds to a markdown heading in the GitHub style.

\hypertarget{summary}{%
\subsection{Pull Request Summary}\label{wfpullrequestsummary}}

This pull request introduces a CodeQL workflow to enhance the security
analysis of this repository.

\hypertarget{what-is-codeql}{%
\subsection{What is CodeQL}\label{what-is-codeql}}

CodeQL is a static analysis tool that helps identify and mitigate
security vulnerabilities. It is primarily intra-function but does
provide some support for inter-function analysis. By integrating CodeQL
into a GitHub Actions workflow, it can proactively identify and address
potential issues before they become security threats.

For more information on CodeQL and how to interpret its results, refer
to the GitHub documentation and the CodeQL documentation
(https://codeql.github.com/ and https://codeql.github.com/docs/).

\hypertarget{what-this-pr-does}{%
\subsection{What this PR does}\label{what-this-pr-does}}

We added a new CodeQL workflow file (.github/workflows/codeql.yml) that

\begin{itemize}[leftmargin=*]
\item
  Runs on every pull request (functionality to run on every push to main
  branches is included as a comment for convenience).
\item
  Runs daily.
\item
  Excludes queries with a high false positive rate or low-severity
  findings.
\item
  Does not display results for git submodules, focusing only on our own
  codebase.
\end{itemize}

\hypertarget{validation}{%
\subsection{Validation}\label{validation}}

To validate the functionality of this workflow, we have run several test
scans on the codebase and reviewed the results. The workflow
successfully compiles the project, identifies issues, and provides
actionable insights while reducing noise by excluding certain queries
and third-party code.

\hypertarget{using-the-workflow-results}{%
\subsection{Using the workflow
results}\label{using-the-workflow-results}}

If this pull request is merged, the CodeQL workflow will be
automatically run on every push to the main branch and on every pull
request to the main branch. To view the results of these code scans,
follow these steps:

Under the repository name, click on the Security tab. In the left
sidebar, click Code scanning alerts.

\subsection{Is this a good idea?}
We are researchers at Purdue University in the USA. We are studying the potential benefits and costs of using CodeQL on open-source repositories of embedded software.

We wrote up a report of our findings so far. The TL;DR is ``CodeQL outperforms the other freely-available static analysis tools, with fairly low false positive rates and lots of real defects''. You can read about the report here: \url{https://arxiv.org/abs/2310.00205}.

\hypertarget{review-of-engineering-hazards}{%
\subsection{Review of engineering
hazards}\label{review-of-engineering-hazards}}

License: see the license at
https://github.com/github/codeql-cli-binaries/blob/main/LICENSE.md:

Here's what you may also do with the Software, but only with an Open
Source Codebase and subject to the License Restrictions provisions
below:
\begin{itemize}
\item  Perform analysis on the Open Source Codebase.
\item If the Open Source Codebase is hosted and maintained on GitHub.com,
generate CodeQL databases for or during automated analysis, CI, or CD.
\end{itemize}

False positives: We find that around 20\% of errors are false positives,
but that these FPs are polarized and only a few rules contribute to most
FPs. We find that the top rules contributing to FPs are:
cpp/uninitialized-local, cpp/missing-check-scanf,
cpp/suspicious-pointer-scaling, cpp/unbounded-write,
cpp/constant-comparison, and cpp/inconsistent-null-check. Adding a
filter to filter out certain rules that contribute to a high FP rate can
be done simply in the workflow file.

%% file: tables/rulewisefps.tex
\begin{table*}[p]
\centering
\tiny
\caption{The numbers of true positives, false positives, total reports, and repositories where defects are reported by \codeql for each \codeql query on a sample of \numreposfortpfp repositories in our dataset.}
\label{tab:DetailedCodeQLQueryPerformance}
\begin{tabular}{@{}lccccc@{}}
\toprule
\textbf{Query}                                   & \textbf{\#TP} &  \textbf{\#FP} & \textbf{\#Results} & \textbf{\#Repo} \\ \midrule
cpp/alloca-in-loop                               & 3   & 0   & 3   & 2  \\
cpp/assignment-does-not-return-this              & 2   & 0   & 2   & 1  \\
cpp/assign-where-compare-meant                   & 0   & 2   & 2   & 2  \\
cpp/bad-addition-overflow-check                  & 1   & 0   & 1   & 1  \\
cpp/badly-bounded-write                          & 1   & 0   & 1   & 1  \\
cpp/bad-strncpy-size                             & 5   & 4   & 9   & 5  \\
cpp/cleartext-transmission                       & 1   & 0   & 1   & 1  \\
cpp/comma-before-misleading-indentation          & 0   & 1   & 1   & 1  \\
cpp/comparison-of-identical-expressions          & 1   & 0   & 1   & 1  \\
cpp/comparison-precedence                        & 6   & 0   & 6   & 1  \\
cpp/constant-comparison                          & 197 & 8   & 205 & 41 \\
cpp/double-free                                  & 0   & 1   & 1   & 1  \\
cpp/duplicate-include-guard                      & 27  & 0   & 27  & 6  \\
cpp/futile-params                                & 4   & 0   & 4   & 3  \\
cpp/implicit-bitfield-downcast                   & 1   & 0   & 1   & 1  \\
cpp/implicit-function-declaration                & 18  & 1   & 19  & 5  \\
cpp/incomplete-parity-check                      & 3   & 0   & 3   & 2  \\
cpp/inconsistent-loop-direction                  & 1   & 0   & 1   & 1  \\
cpp/inconsistent-null-check                      & 181 & 2   & 183 & 35 \\
cpp/incorrect-allocation-error-handling          & 15  & 0   & 15  & 1  \\
cpp/incorrect-not-operator-usage                 & 6   & 1   & 7   & 4  \\
cpp/integer-used-for-enum                        & 1   & 0   & 1   & 1  \\
cpp/logical-operator-applied-to-flag             & 1   & 6   & 7   & 3  \\
cpp/lossy-function-result-cast                   & 28  & 0   & 28  & 3  \\
cpp/memset-may-be-deleted                        & 2   & 0   & 2   & 2  \\
cpp/missing-case-in-switch                       & 1   & 0   & 1   & 1  \\
cpp/missing-check-scanf                          & 58  & 41  & 99  & 18 \\
cpp/missing-return                               & 2   & 0   & 2   & 2  \\
cpp/mistyped-function-arguments                  & 0   & 23  & 23  & 4  \\
cpp/nested-loops-with-same-variable              & 8   & 0   & 8   & 5  \\
cpp/new-array-delete-mismatch                    & 1   & 0   & 1   & 1  \\
cpp/non-constant-format                          & 2   & 0   & 2   & 1  \\
cpp/non-https-url                                & 6   & 0   & 6   & 1  \\
cpp/non-member-const-no-effect                   & 3   & 0   & 3   & 1  \\
cpp/offset-use-before-range-check                & 19  & 4   & 23  & 15 \\
cpp/overflow-destination                         & 3   & 3   & 6   & 4  \\
cpp/overflowing-snprintf                         & 15  & 0   & 15  & 3  \\
cpp/overrunning-write                            & 9   & 7   & 16  & 8  \\
cpp/overrunning-write-with-float                 & 4   & 0   & 4   & 2  \\
cpp/pointer-overflow-check                       & 1   & 0   & 1   & 1  \\
cpp/redefined-default-parameter                  & 3   & 0   & 3   & 1  \\
cpp/redundant-null-check-simple                  & 2   & 0   & 2   & 2  \\
cpp/resource-not-released-in-destructor          & 3   & 0   & 3   & 1  \\
cpp/rule-of-two                                  & 5   & 0   & 5   & 1  \\
cpp/signed-overflow-check                        & 3   & 0   & 3   & 3  \\
cpp/stack-address-escape                         & 97  & 67  & 164 & 29 \\
cpp/static-buffer-overflow                       & 1   & 0   & 1   & 1  \\
cpp/string-copy-return-value-as-boolean          & 4   & 0   & 4   & 1  \\
cpp/suspicious-add-sizeof                        & 0   & 1   & 1   & 1  \\
cpp/suspicious-allocation-size                   & 0   & 3   & 3   & 2  \\
cpp/suspicious-pointer-scaling                   & 0   & 19  & 19  & 9  \\
cpp/suspicious-pointer-scaling-void              & 1   & 34  & 35  & 5  \\
cpp/suspicious-sizeof                            & 5   & 2   & 7   & 6  \\
cpp/toctou-race-condition                        & 13  & 0   & 13  & 8  \\
cpp/too-few-arguments                            & 1   & 0   & 1   & 1  \\
cpp/unbounded-write                              & 27  & 29  & 56  & 24 \\
cpp/uncontrolled-allocation-size                 & 39  & 11  & 50  & 13 \\
cpp/uncontrolled-arithmetic                      & 3   & 7   & 10  & 3  \\
cpp/uncontrolled-process-operation               & 6   & 0   & 6   & 4  \\
cpp/uninitialized-local                          & 30  & 251 & 281 & 41 \\
cpp/unsafe-strcat                                & 7   & 5   & 12  & 6  \\
cpp/unsigned-comparison-zero                     & 18  & 0   & 18  & 8  \\
cpp/unsigned-difference-expression-compared-zero & 11  & 0   & 11  & 5  \\
cpp/unterminated-variadic-call                   & 1   & 0   & 1   & 1  \\
cpp/use-after-free                               & 6   & 1   & 7   & 5  \\
cpp/use-in-own-initializer                       & 1   & 0   & 1   & 1  \\
cpp/useless-expression                           & 59  & 2   & 61  & 13 \\
cpp/user-controlled-bypass                       & 2   & 2   & 4   & 2  \\
cpp/virtual-call-in-constructor                  & 6   & 0   & 6   & 1  \\
cpp/weak-cryptographic-algorithm                 & 10  & 0   & 10  & 2  \\
cpp/wrong-number-format-arguments                & 3   & 0   & 3   & 2  \\
cpp/wrong-type-format-argument                   & 35  & 0   & 35  & 5  \\
\midrule
\textbf{Total}                                   & \textbf{1039}    & \textbf{538}  & \textbf{1577}       & \textbf{}       \\ \bottomrule
\end{tabular}
\end{table*}